\documentclass[aps,prd,twocolumn,preprintnumbers,
superscriptaddress,showpacs,floatfix]{revtex4}

\usepackage{bm}
\usepackage{epsfig}
\usepackage{graphics}

\newcommand{\eeq}{\end{equation}}
\newcommand{\beq}{\begin{equation}}
\newcommand{\ba}{\begin{array}}
\newcommand{\ea}{\end{array}}
\newcommand{\bea}{\begin{eqnarray}}
\newcommand{\eea}{\end{eqnarray}}
\newcommand{\ftn}{\footnotesize}

\newcommand{\ssz}{\scriptsize}

\newcommand{\etal}{{\it et al.\/}}
\newcommand{\tr}{{\mbox{\sf\ssz T}}}

\newcommand{\ns}{\ensuremath{n_{\rm s}}}

\def\rsym{\leavevmode\hbox{\scriptsize $R$}}
\def\rsm{\leavevmode\hbox{R}}
\newcommand\vev[1]{\langle {#1} \rangle}
\def\lf{\left(}
\def\rg{\right)}
\def\lfa{\left|}
\def\rga{\right|}
\def\llgm{\left\lgroup\matrix}
\def\rrgm{\right\rgroup}

\newcommand{\Gr}{\ensuremath{\widetilde{G}}}
\newcommand{\Trha}{\ensuremath{T_{\rm 1rh}}}
\newcommand{\Trhb}{\ensuremath{T_{\rm 2rh}}}
\newcommand{\Vhi}{\ensuremath{V_{\rm HI}}}
\newcommand{\Vhio}{\ensuremath{V_{\rm HI0}}}
\newcommand{\Vpq}{\ensuremath{V_{\rm PQ}}}
\newcommand{\Vpqo}{\ensuremath{V_{\rm PQ0}}}
\newcommand{\ck}{{\ensuremath\mbox{\sl a}}}
\newcommand{\cks}{{\mbox{\ftn\sl a}}}
\newcommand{\ckss}{{\mbox{\ssz\sl a}}}
\newcommand{\mP}{\ensuremath{m_{\rm P}}}
\newcommand{\ld}{\ensuremath{\lambda}}
\newcommand{\ka}{\ensuremath{\kappa_a}}
\newcommand{\kp}{\ensuremath{\kappa}}
\newcommand{\GeV}{{\mbox{\rm GeV}}}

\newcommand{\sg}{\ensuremath{\sigma}}

\newcommand{\sgap}{\ensuremath{\sigma_{1\rm +}}}
\newcommand{\sgam}{\ensuremath{\sigma_{1\rm -}}}

\newcommand{\spl}{\ensuremath{s_{\rm +}}}
\newcommand{\sm}{\ensuremath{s_{\rm -}}}
\newcommand{\Ap}{\ensuremath{A_{\rm +}}}
\newcommand{\Am}{\ensuremath{A_{\rm -}}}
\newcommand{\Vp}{\ensuremath{U_{\rm +}}}
\newcommand{\Vm}{\ensuremath{U_{\rm -}}}
\newcommand{\Np}{\ensuremath{N_{\rm +}}}
\newcommand{\Nm}{\ensuremath{N_{\rm -}}}
\newcommand{\msp}{\ensuremath{m^2_{+}}}
\newcommand{\msm}{\ensuremath{m^2_{-}}}
\newcommand{\mspm}{\ensuremath{m^2_{\pm}}}
\def\Ka{K\"{a}hler potential}

\renewcommand{\theparagraph}{{\bf\alph{paragraph}}}

\begin{document}

\preprint{UT-STPD-2/10}

\title{F-Term Hybrid Inflation Followed by a Peccei-Quinn Phase
Transition}
%Based on Renormalizable Terms
\author{G. Lazarides}
\email{lazaride@eng.auth.gr} \affiliation{Physics Division, School
of Technology, Aristotle University of Thessaloniki, Thessaloniki
54124, GREECE}
\author{C. Pallis}
\email{kpallis@gen.auth.gr} \affiliation{Department of Physics,
University of Cyprus, P.O. Box 20537, Nicosia 1678, CYPRUS}

%\date{\today}

\begin{abstract}
We consider a cosmological set-up, based on renormalizable
superpotential terms, in which a superheavy scale F-term hybrid
inflation is followed by a Peccei-Quinn phase transition,
resolving the strong CP and $\mu$ problems of the minimal
supersymmetric standard model. We show that the field which
triggers the Peccei-Quinn phase transition can remain after
inflation well above the Peccei-Quinn scale thanks to (i) its
participation in the supergravity and logarithmic corrections
during the inflationary stage and (ii) the high reheat temperature
after the same period. As a consequence, its presence influences
drastically the inflationary dynamics and the universe suffers a
second period of reheating after the Peccei-Quinn phase
transition. Confronting our inflationary predictions with the
current observational data, we find that, for about the central
value of the spectral index, the grand unification scale can be
identified with its supersymmetric value for the relevant coupling
constant $\kappa\simeq0.002$ and, more or less, natural values,
$\pm(0.01-0.1)$, for the remaining parameters. On the other hand,
the final reheat temeperature after the Peccei-Quinn phase
transition turns out to be low enough so as the gravitino problem
is avoided.
\end{abstract}

 \maketitle

\section{Introduction}

One of the most natural and well-motivated inflationary model is
the \emph{supersymmetric} (SUSY) \emph{F-term hybrid inflation}
(FHI) \cite{hybrid,susyhybrid}. It is realized at (or close to)
the SUSY \emph{Grand Unified Theory} (GUT) scale $M_{\rm
GUT}\simeq2.86\times10^{16}~{\rm GeV}$ and can be easily linked to
extensions \cite{lectures} of the \emph{Minimal Supersymmetric
Standard Model} (MSSM) which provide solutions to a number of
problems of MSSM. Namely, the $\mu$-problem of MSSM can be solved
via a direct coupling of the inflaton to Higgs superfields
\cite{dvali} or via a \emph{Peccei-Quinn} (PQ) symmetry
\cite{rsym} which also solves the strong CP problem \cite{pq}.
Also baryon number conservation can be an automatic consequence
\cite{dvali} of a R symmetry and the baryon asymmetry of the
universe can be generated via leptogenesis which takes place
\cite{lept} through the out-of-equilibrium decay of the inflaton's
decay products.

The aforementioned resolution of the $\mu$-problem of MSSM via a
PQ symmetry can be achieved \cite{rsym} by considering
non-renormalizable superpotential  terms involving additional
singlets which develop \emph{vacuum expectation values} (VEVs) of
the order of the PQ symmetry breaking scale. The emerging
potential of these singlets has a local minimum which is separated
from the global PQ minimum by a sizable potential barrier
preventing a successful transition from the trivial to the PQ
vacuum. As one can show \cite{jean}, by considering the one-loop
temperature corrections \cite{jackiw} to the scalar potential,
this situation persists at all cosmic temperatures after
reheating. It is, thus, obligatory to assume that, after the
termination of FHI, the system emerges with the appropriate
combination of initial conditions so that it is led
\cite{curvaton} to the PQ vacuum. Stated differently, the PQ
symmetry is to be broken during or before FHI. As a consequence,
tight upper bounds on the inflationary scale have to be imposed so
that the isocurvature fluctuations \cite{isoR} of the axion are
consistent with the observational bounds -- this restriction,
though, can be alleviated when cosmic strings are formed
\cite{lyth}.

This latter complication can be avoided, if the \emph{PQ phase
transition} (PQPT) takes place  after the end of FHI. This
possibility can be realized adopting \cite{goto} only
renormalizable superpotential terms similar to those which lead to
FHI. The $\mu$ parameter of MSSM can also be generated from the PQ
scale as in Ref.~\cite{rsym}. On the other hand, this scheme may
lead \cite{sikivie} to disastrous domain walls which, however, can
be avoided \cite{georgi, cdmon} by introducing extra matter
superfields without jeopardizing the unification of the MSSM gauge
coupling constants. Moreover, the structure of the superpotential
obliges us to consider extra couplings between the inflaton field
of FHI and the inflaton-like field of PQPT in the K\"alher
potential. As a bonus, these couplings assist us to reconcile the
results on the (scalar) spectral index $n_{\rm s}$ of FHI and the
recent seven-year results \cite{wmap3} from the \emph{Wilkinson
microwave anisotropy probe} (WMAP7) satellite.

Indeed, it is well-known that the realization of FHI within
minimal \emph{Supergravity} (SUGRA) leads to $\ns$ which is just
marginally consistent with the fitting of the WMAP7 data by the
standard power-law cosmological model with \emph{cold dark matter
and a cosmological constant} ($\Lambda$CDM). One possible
resolution (for other proposals, see Refs.~\cite{battye, mhi,
hinova, axilleas, news}) of this problem is \cite{gpp, king} the
addition to the K\"ahler potential of a non-minimal quatric term
of the inflaton field with a convenient choice of its sign. As a
consequence, a negative mass term for the inflaton is generated.
In the largest part of the parameter space, the inflationary
potential acquires a local maximum and minimum. Then, FHI of the
hilltop type \cite{lofti} can occur as the inflaton rolls from
this maximum down to smaller values. Therefore, $\ns$ can become
consistent with data, but only at the cost of an extra
indispensable mild tuning \cite{gpp} of the initial conditions.
Another possible complication is that the system gets trapped near
the minimum of the inflationary potential and, consequently, no
FHI takes place.

Within our proposal, a negative mass term for the inflaton is
generated too, which however depends on the various coefficients
(of the order $0.01-0.1$) involved in the \Ka\ and not exclusively
on the quatric term of the inflaton field. As a consequence,
sufficiently low $\ns$ can be achieved with either sign of the
terms in the \Ka\ and the inflationary potential remains monotonic
during the period of FHI. Therefore, complications due to the
appearance of maxima and minima along the inflationary path can be
eluded. Another by-product of our proposal is that the reheat
temperature after PQPT, calculated consistently with the model
interactions, turns out to be low enough so that the gravitino
($\Gr$) constraint \cite{gravitino, kohri} and the potential
problem of topological defects \cite{kibble} of FHI can be
significantly relaxed since an entropy release takes place after
PQPT. On the other hand, the reheat temperature is high enough so
that non-perturbative electroweak sphalerons are operative and,
consequently, thermal \cite{lepto} or non-thermal \cite{ntlepto}
leptogenesis can, in principle, work.

Below, we present the basic ingredients of our model
(Sec.~\ref{fhim}) and describe the inflationary potential and
dynamics (Secs.~\ref{fhi1} and \ref{fhi2}). We then exhibit the
constraints imposed on our cosmological set-up (Sec.~\ref{cont}).
We end up with our numerical results (Sec.~\ref{num}) and our
conclusions (Sec.~\ref{con}).

\section{Model Description}\label{fhim}

\subsection{The general set-up}\label{model}

In order to explore our scenario, we adopt the left-right
symmetric gauge group $G_{\rm LR} ={\rm SU(3)_c}\times$ ${\rm
SU(2)_L\times SU(2)_R \times U(1)}_{B-L}$. In our scheme, $G_{\rm
LR}$ can be broken down to the standard model gauge group $G_{\rm
SM}$ at a scale close to the SUSY GUT scale $M_{\rm GUT}$ through
the VEVs acquired by a conjugate pair of SU$(2)_{\rm R}$ doublet
left-handed Higgs superfields, $\bar\Phi$ and $\Phi$, with
$B-L=-1,~1$ respectively. As a consequence, no cosmic strings are
produced \cite{kibble, trotta} in this realization of standard FHI
and, therefore, we are not obliged to impose extra restrictions on
the parameters -- as e.g. in Ref.~\cite{mairi}.

The model possesses also three global U(1) symmetries. Namely, a
(color) anomalous PQ symmetry ${\rm U(1)}_{\rm PQ}$, an anomalous
R symmetry ${\rm U(1)_{\rm \rsym}}$, and the baryon number
symmetry ${\rm U(1)}_B$. Note that global continuous symmetries
can effectively arise \cite{laz1} from the rich discrete symmetry
groups encountered in many compactified string theories (see e.g.
Ref.~\cite{laz2}). The PQ symmetry ${\rm U(1)}_{\rm PQ}$ can be
spontaneously broken at the PQ breaking scale
$f_a\sim(10^{10}-10^{12})~{\rm GeV}$ (which coincides with the
axion decay constant -- for a review see Ref.~\cite{kim}) via the
VEVs acquired by two $G_{\rm LR}$ singlet left-handed superfields
$\bar Q$ and $Q$.

\begin{table}[!t]
\caption{Superfield Content of the Model}
\begin{tabular}{c@{\hspace{0.4cm}}c@{\hspace{0.4cm}}c@{\hspace{0.4cm}}c@{\hspace{0.4cm}}c@{\hspace{0.4cm}}c}
\toprule {Superfields}&{Representations}&\multicolumn{4}{c}{Global
Symmetries}
\\ \multicolumn{1}{c}{}&{under $G_{\rm LR}$}
&\rsm\ &{PQ} &{$B$}&{$D$}
\\\colrule
\multicolumn{6}{c}{Matter Fields}\\\colrule
{$l_i$} &{$({\bf 1, 2, 1, -1})$}& $0$ & $-2$ &$0$&$0$
 \\
{$l^c_i$} & {$({\bf 1, 1, 2, 1})$} &$2$&{$0$}&{$0$}&$0$
\\
{$q_i$} &{$({\bf 3, 2, 1, 1/3})$}& $1$ & $-1$ &$1/3$&$0$
 \\
{$q^c_i$} & {$({\bf \bar 3, 1, 2,-1/3})$} &$1$ &{$-1$}&$-1/3$&$0$
\\ \colrule
\multicolumn{6}{c}{Higgs Fields}
\\\colrule
{$S$} & {$({\bf 1, 1, 1, 0})$}&$4$ &$0$ &$0$&$0$ \\
{$\bar \Phi$}&$({\bf 1, 1, 2, -1})$&{$0$}&{$0$}&{$0$}&$0$\\
{$\Phi$} &{$({\bf 1, 1, 2, 1})$}&{$0$}&{$0$} & {$0$}&$0$ \\
\colrule
{$P$} & {$({\bf 1, 1, 1, 0})$}&$4$ &$0$ &$0$&$0$\\
{$\bar Q$}&$({\bf 1, 1, 1, 0})$& {$0$}&{$-2$}&{$0$}&$0$\\
{$Q$} &{$({\bf 1, 1, 1, 0})$}& {$0$}&{$2$} & {$0$}&$0$ \\ \colrule
{$h$} & {$({\bf 1, 2, 2, 0})$}&$2$ &$2$ &$0$&$0$
\\ \colrule
\multicolumn{6}{c}{Extra Matter Fields} \\ \colrule
$\bar D_{\rm a}$&{$({\bf \bar 3, 1, 1, 2/3})$}& $2$ & $1$
&$0$&$-1$\\
$D_{\rm a}$&{$({\bf 3, 1, 1, -2/3})$}& $2$ & $1$ &$0$&$1$\\
$H_{\rm a}$&$({\bf 1, 2, 2, 0})$ & $2$ & $1$ &$0$&$0$
\\ \botrule
\end{tabular}
\label{tabLR}
\end{table}

The part of the superpotential which is relevant for the breaking
of the group $G_{\rm LR}\times {\rm U(1)}_{\rm PQ}$ is
\beq\label{Whi} W= \kappa S\left(\bar\Phi\Phi-M^2\right)+\kappa_a
P(\bar Q Q-f^2_a/4)+ \lambda S\bar Q Q, \eeq
%%%%%%%%%%%%%%%%%%%%%%%
where $S$, $P$ are $G_{\rm LR}$ singlet left-handed superfields
which trigger the breaking of $G_{\rm LR}$ and ${\rm U(1)}_{\rm
PQ}$ respectively and the parameters $\kappa$, $\kappa_a$, $M\sim
M_{\rm GUT}$, and $f_a$ are made positive by field redefinitions.
For simplicity, we restrict our analysis to real $\lambda$'s. The
superfield $P$ can be regarded as the linear combination of the
$G_{\rm LR}$ singlets with PQ and R charge equal to 0 and 4
respectively that does not couple to $\bar\Phi\Phi$ -- cf.
Ref.~\cite{tetradis}. In this basis, the most general K\"ahler
potential of our model includes interference terms of $S$ and $P$
even at the quadratic level -- contrary to the choice opted in
Ref.~\cite{twin}. Namely, we adopt the following K\"ahler
potential
\bea K&=&|S|^2+|P|^2+\ck (S P^*+ S^*P)\nonumber\\&&+\
b{|S|^4\over4\mP^2}+c{|P|^4\over4\mP^2}+
d{|S|^2|P|^2\over\mP^2}\nonumber\\&&+\ { e|S|^2+
f|P|^2\over2\mP^2}\lf S P^*+ S^*P\rg\nonumber\\&&+\
{g\over4\mP^2}\left[\lf S P^*\rg^2+ \lf
S^*P\rg^2\right]\nonumber\\&& +\ |\Phi|^2+|\bar\Phi|^2+|Q|^2+|\bar
Q|^2 \nonumber\\&& +\ k (\Phi\bar \Phi+\Phi^*\bar \Phi^*)  + l
(Q\bar Q+Q^*\bar Q^*) +\ \cdots,\label{minK}\eea
where all the coefficients $\ck,b,c,d,e,f, g, k,$ and $l$ are
taken, for simplicity, real although some of them can be, in
general, complex. The ellipsis represents higher order terms
involving the waterfall fields ($\Phi$, $\bar \Phi$, $Q$, and
$\bar Q$) and $S$ and $P$. We can neglect these terms since they
have negligible impact on the SUSY vacuum and are irrelevant along
the inflationary path -- see below.

The usual superpotential terms of MSSM can be derived from the
following superpotential:
\beq\label{Wmu} W_{\rm m} =\lambda_\mu {\bar Q^2h^2\over 2m_{\rm
P}}+y_{\nu ij} {\bar \Phi l^c_i \bar \Phi l^c_j\over m_{\rm
P}}+y_{l ij} l_ihl^c_j  + y_{q ij} q_ihq^c_j \eeq
with $m_{\rm P}\simeq 2.44\times 10^{18}~{\rm GeV}$ being the
reduced Planck scale. Here, the $i$th generation ${\rm SU(2)_L}$
doublet left-handed quark and lepton superfields are denoted by
$q_i$ and $l_i$ respectively, whereas the ${\rm SU(2)_R}$ doublet
antiquark and antilepton superfields by $q^c_i$ and $l^c_i$
respectively. The electroweak Higgs superfields are contained in a
${\rm SU(2)_L}\times{\rm SU(2)_R}$ bidoublet Higgs superfield $h$.
The first term in the \emph{right-hand side} (RHS) of
Eq.~(\ref{Wmu}) generates the $\mu$ term of MSSM via the PQ
breaking scale (see below), while the second term generates
intermediate scale masses for the right-handed neutrinos and,
thus, seesaw masses \cite{susyhybrid} for the light neutrinos.

The representations under $G_{\rm LR}$ and the charges under the
global symmetries of the various matter and Higgs superfields
contained in this model are presented in Table~\ref{tabLR}, which
also contains the extra matter superfields required for evading
the domain wall problem associated with the PQPT together with a
new imposed global symmetry ${\rm U(1)}_D$ -- see
Sec.~\ref{walls}.

\subsection{The cosmological scenario}\label{inflation12}

The F--term SUGRA scalar potential obtained from $W$ in
Eq.~(\ref{Whi}) and $K$ in Eq.~(\ref{minK}) can be found by
applying the well-known formula (see e.g. Ref.~\cite{hybrid})
\begin{equation}
V_{\rm SUGRA}=e^{K/\mP^2}\left(F_{i^*}^*\lf K_{,ji^*}\rg^{-1}
F_j-3\frac{\vert W\vert^2}{\mP^2}\right), \label{sugra}
\end{equation}
where $F_i=W_{,i} +K_{,i}W/\mP^2$ and a subscript $,i~[,i^*]$
denotes derivation \emph{with respect to} (w.r.t.) the complex
scalar field $i~[i\,^{*}]$. In the limit where $\mP$ tends to
infinity, we can obtain the SUSY limit, $V_{\rm F}$, of $V_{\rm
SUGRA}$ in Eq.~(\ref{sugra}) which turns out to be
\bea \label{VF} V_{\rm F} &\simeq&
\left|\kappa\left(\bar\Phi\Phi-M^2\right)+\lambda  \bar Q
Q\right|^2/(1-\ck^2)\nonumber\\
&&+\ \kappa_a^2\left|\bar Q Q-f^2_a/4\right|^2/(1-\ck^2)\nonumber\\
&&+\ \kappa^2|S|^2\left(|\bar\Phi|^2+ |\Phi|^2\right)\nonumber\\
&&+\ \left|\lambda  S+\kappa_a P\right|^2\left(|\bar Q|^2+
|Q|^2\right)\nonumber\\
&&-\ \Big[{\ck\kappa_a}\lf \bar Q^*
Q^*-f^2_a/4\rg\big[\kappa\lf\bar\Phi\Phi-M^2\right)\nonumber\\
&&+\ \lambda \bar Q Q\big]/(1-\ck^2)+ \mbox{complex
conjugate}\Big],\eea
where the complex scalar fields which belong to the SM singlet
components of the superfields are denoted by the same symbol and
we take into account that $M\gg f_a$. Note that although the
mixing terms in the first line of Eq.~(\ref{minK}) have an impact
on Eq.~(\ref{VF}), the corresponding terms in the last line of
this equation do not contribute since they are holomorphic or
anti-holomorphic in the field variables. Also the terms in the
ellipsis in Eq.~(\ref{minK}) have no appreciable contribution to
the RHS of Eq.~(\ref{VF}) since they are suppressed by powers of
$\mP$.

The D--term contribution vanishes along the direction
\beq |\bar\Phi|=|\Phi|.\label{Dflat}\eeq
From the potential in Eq.~(\ref{VF}) and taking into account that
$M\gg f_a$, we find that the SUSY vacuum lies at
\beq\langle S\rangle=0,~
|\langle\bar\Phi\rangle|=|\langle\Phi\rangle|\simeq M,~ \langle
P\rangle=0,~\mbox{and}~|\langle\phi_{Q}\rangle|=f_a,\label{vevs}
\eeq
where we have introduced the canonically normalized scalar field
$\phi_{Q}=2Q=2\bar Q$. Note that, since the sum of the arguments
of $\vev{\bar Q}$, $\vev{Q}$ must be $0$, $\bar Q$ and $Q$ can be
brought to the real axis by an appropriate PQ transformation. As a
consequence of Eqs.~(\ref{vevs}), $W$ leads to a spontaneous
breaking of $G_{\rm LR}$ and ${\rm U}_{\rm PQ}(1)$. The same
superpotential $W$ gives also rise to a stage of FHI and a PQPT,
which is highlighted in the following.

The potential in Eq.~(\ref{VF}) possesses a D-- and F--flat
direction at
\beq \bar \Phi=\Phi=0~\mbox{and}~\bar Q= Q=0\label{FDflat}\eeq
with a constant potential energy density
\beq \Vhio\simeq\frac{\kappa^2M^4}{1-\ck^2}~~(\mbox{since}~~M\gg
f_a).\label{V1}\eeq
The direction in Eq.~(\ref{FDflat}) can be used as the
inflationary trajectory since it corresponds to a classically flat
valley of minima for
\beq\label{flat}{\sf
(a)}~|S|>{M\over\sqrt{1-\ck^2}}~~\mbox{and}~~{\sf (b)}~|\sigma_a|>
{\sqrt{\kappa(\lambda-\ck\ka)\over1-\ck^2}}M,\eeq
where we have defined $\sigma_a=\lambda S+\kappa_aP$.

\begin{table}[!t]
\caption{The mass spectrum of the model along the inflationary
trajectory of Eqs.~(\ref{FDflat}) and (\ref{flat}).}
\begin{tabular}{c|@{\hspace{0.1cm}}c@{\hspace{0.1cm}}|@{\hspace{0.1cm}} c}\toprule
{Superfields} &{Fields} & {Mass Squared}\\
{of origin} & {}&{}\\\colrule
\multicolumn{3}{c}{Bosons}\\ \colrule
$\bar\Phi$, $\Phi$ & $4$ complex scalars &
$\kp^2\lf\lfa S\rga^2\pm \frac{M^2}{1-\ckss^2}\rg$ \\
$\bar Q$, $Q$ & 2 complex scalars & $\lfa \sigma_a\rga^2
\pm\frac{\kappa(\lambda-\ckss\ka)M^2}{1-\ckss^2}$\\\colrule
\multicolumn{3}{c}{Fermions}\\ \colrule
$\bar\Phi$, $\Phi$ & $2$ Dirac spinors & $\kappa^2|S|^2$\\
$\bar Q$, $Q$ & 1 Dirac spinors & $|\sigma_a|^2$ \\
\botrule
\end{tabular}
\label{tab1}
\end{table}

For relatively large $\lambda$'s, the reheating temperature after
FHI turns out to be rather high, leading to troubles with the
$\Gr$ abundance \cite{gravitino, kohri} -- as usually in SUSY
cosmological models. However, we can verify that a D- and F-flat
direction appears at
\beq S=0,~\bar \Phi=\Phi=M,~\mbox{and}~\bar Q= Q=0
\label{PQflat}\eeq
with potential energy density
\beq \Vpqo=\kappa_a^2f_a^4/16.\label{V2}\eeq
Since $\Vpqo\ll\Vhio$, $\Vpqo$ can temporally dominate over
radiation after the end of FHI and, if $|P|\geq f_a$, drive a PQPT
-- see Sec.~\ref{fhi2}. A subsequent second episode of reheating
can dilute sufficiently the unwanted $\Gr$ concentration -- see
Appendix~\ref{Rhg}. At the end of PQPT, the fields $P$, $\bar Q$,
and $Q$ acquire their VEVs in Eq.~(\ref{vevs}). The $\mu$ term of
the MSSM is generated via the first term in the RHS of
Eq.~(\ref{Wmu}) with $|\mu|\sim\lambda_\mu\left|\vev{\bar
Q}\right|^2/m_{\rm P}$, which is of the right magnitude if
$\left|\vev{\bar Q}\right|=f_a/2\simeq 5\times10^{11}~{\rm GeV}$
and $\lambda_\mu\sim(0.001-0.01)$.

The cosmological scenario above can be attained if we ensure that,
at the end of FHI, the fields $\bar \Phi$ and $\Phi$ acquire their
VEVs in Eq.~(\ref{vevs}), while the fields $\bar Q$ and $Q$ remain
equal to zero. To quantify this crucial requirement,  we construct
the mass spectrum of the theory along the inflationary path of
Eqs.~(\ref{FDflat}) and (\ref{flat}). We summarize our results in
Table~\ref{tab1}. We see that the mass matrices of the scalar
components of the $\bar \Phi$, $\Phi$ and $\bar Q$, $Q$
superfields develop a negative eigenvalue as $|S|$ and
$|\sigma_a|$ cross below their critical values -- i.e. their lower
bounds in Eq.~(\ref{flat}). However, if the tachyonic instability
of the $\bar \Phi-\Phi$ system occurs first, $\bar \Phi$ and
$\Phi$ start evolving towards their VEVs, whereas $\bar Q$ and $Q$
continue to be confined to zero.

\subsection{Evading the domain-wall problem}\label{walls}

It should be mentioned that instanton and soft SUSY breaking
effects explicitly break ${\rm U(1)_{\rm \rsym}}\times {\rm
U(1)_{\rm PQ}}$ to a discrete subgroup. Spontaneous breaking of
this subgroup at the PQPT by $\vev{\bar Q}$ and $\vev{Q}$ can
lead \cite{sikivie} to a disastrous domain wall
production since this transition occurs after FHI. In order to
avoid this problem, we must introduce some extra matter fields
(see Table~\ref{tabLR}) following Ref.~\cite{georgi} (see also
Ref.~\cite{cdmon}, where the set-up is similar to ours). Namely,
we introduce $n$ pairs of left-handed superfields $\bar D_{\rm a}$
and $D_{\rm a}$ (${\rm a} = 1, ..., n$) which are ${\rm SU(3)_c}$
antitriplets and triplets respectively with R charge equal to 2
and PQ charge equal to 1. However, these fields acquire
intermediate scale masses after the PQ breaking, which could
prevent the unification of the MSSM gauge coupling constants. To
restore gauge unification, we include an equal number of ${\rm
SU(2)_L}\times{\rm SU(2)_R}$ bidoublet superfields $H_{\rm a}$
with PQ and R charges equal to those of $\bar D_{\rm a}$ and
$D_{\rm a}$. In accordance with all the imposed symmetries, we can
give intermediate scale masses to $\bar D_{\rm a}-D_{\rm a}$ and
$H_{\rm a}$ through the superpotential terms
\beq\label{Wdw} W_{\rm dw} =\lambda_{D\rm a}\bar Q \bar D_{\rm a}
D_{\rm a}+\lambda_{H\rm a}\bar Q H_{\rm a}^2. \eeq
Here, we chose a basis in the $\bar D_{\rm a}-D_{\rm a}$ and
$H_{\rm a}$ space where the coupling constant matrices
$\lambda_{D\rm a}$ and $\lambda_{H\rm a}$ are diagonal. Note that
the full superpotential is invariant under a new global ${\rm
U(1)}_D$ symmetry -- see Table~\ref{tabLR}. During FHI, we take
$\bar D_{\rm a}=D_{\rm a}=H_{\rm a}=0$ and, at the SUSY vacuum, we
have $\vev{\bar D_{\rm a}}=\vev{D_{\rm a}}=\vev{H_{\rm a}}=0$.

The number $n$ can be determined as follows: The explicitly
unbroken subgroup of ${\rm U(1)_{\rm \rsym}}\times {\rm U(1)_{\rm
PQ}}$ can be found, for every $n$, from the solutions of the
system
\beq \nonumber 4\alpha =0~\lf\mbox{\sf mod}~2\pi\rg~\mbox{and}~
-12\alpha +2(n-6)\beta=0~\lf\mbox{\sf mod}~2\pi\rg, \eeq
where $\alpha$ and $\beta$ are the phases of a ${\rm U(1)_{\rm
\rsym}}$ and ${\rm U(1)_{\rm PQ}}$ rotation respectively. Here we
took into account that the \rsm\ charge of $W$ and, thus, of all
the soft SUSY breaking term is 4 and that the sum of the \rsm\
[PQ] charges of the ${\rm SU(3)_c}$ triplets and antitriplets is
$-12$~[$2(n-6)$]. We can induce, therefore, that the explicitly
unbroken subgroup is $Z_4\times Z_{2(n-6)}$. It is then important
to ensure that this subgroup is not spontaneously broken by
$\vev{\bar Q}$ and $\vev{Q}$, i.e.,
\beq\label{sunb} 2\beta=0~\lf\mbox{\sf mod}~2\pi\rg, \eeq
since otherwise cosmologically disastrous domain walls will be
produced at PQPT. This requirement implies that $n$ must be 5 or
7. In both these cases, the subgroup of ${\rm U(1)_{\rm
\rsym}}\times{\rm U(1)_{\rm PQ}}$ left unbroken by instantons and
SUSY breaking coincides with the one left unbroken by $\vev{\bar
Q}$ and $\vev{Q}$ and is a $Z_4\times Z_2$.

It is easy to find the contribution of $\bar D_{\rm a}, D_{\rm
a},$ and $H_{\rm a}$ to the coefficients $b_1,~b_2,$ and $b_3$
controlling \cite{Jones} the one loop evolution of the three gauge
coupling constants $g_1, g_2,$ and $g_3$ within MSSM. It is then
straightforward to prove that, if we assign $B-L=2/3$ and $-2/3$
to $\bar D_{\rm a}$ and $D_{\rm a}$ respectively, the quantities
$b_2-b_1$ and $b_3-b_2$ (which are \cite{Jones} crucial for the
unification of $g_1, g_2,$ and $g_3$) remain unaltered. Therefore,
the inclusion of the extra matter superfields does not disturb the
gauge unification at one loop.

Recapitulating this section, let us present the total
superpotential of our model which is
\beq W_{\rm tot} = W+W_{\rm m}+W_{\rm dw}, \eeq
where $W$, $W_{\rm m}$ and $W_{\rm dw}$ are given in
Eq.~(\ref{Whi}), (\ref{Wmu}), and (\ref{Wdw}) respectively.

\section{The inflationary epoch}\label{fhi1}

\subsection{Structure of the inflationary
potential}\label{fhi1path}

The potential which can drive the inflationary stage of our set-up
has the following general form
\beq\label{Vol} \Vhi=V_{\rm HI0}+V_{\rm HIs}+V_{\rm HIc},\eeq
where $V_{\rm HIs}$ and $V_{\rm HIc}$ represent, respectively,
SUGRA and one-loop radiative corrections to the inflationary
potential, calculated in Sec.~\ref{fhi1s} and \ref{fhi1c}. Let us
note, in passing, that the most important contribution
\cite{sstad} to $\Vhi$ from the soft SUSY breaking terms is
expected to start playing an important role for rather small
$\kappa$'s and so, it remains negligibly small in our set-up due
to the large $\kappa$'s used -- see Sec.~\ref{num}.
%\vspace*{-.6cm}

\subsubsection{Supergravity corrections}\label{fhi1s}

Expanding the F--term SUGRA potential in Eq.~(\ref{sugra}) along
the inflationary trajectory -- see Eq.~(\ref{FDflat}) -- in powers
of $1/\mP$ for $M\gg f_a$, we obtain the following expression:
\vspace*{-0.9cm}
\begin{widetext}
\bea V_{{\rm HIs}}&\simeq& {\Vhio\over (1-\ck^2)m^2_{\rm P}}\Big[
A_1|S|^2 + A_{12}\lf S^*P+PS^*\rg + A_2|P|^2\Big]+{\Vhio\over
4(1-\ck^2)^2m^4_{\rm P}}\Big[ B_1|S|^4 + B_2|P|^4 \nonumber\\&&+\
B_{3} |S|^2|P|^2+\lf B_{4}|S|^2+B_{5}|P|^2\rg \lf
S^*P+PS^*\rg+B_6\lf \lf S^*P\rg^2+\lf P^*S\rg^2\rg
\Big],\label{Vexpan}\eea%\vspace*{-.6cm}
%\lf S^*P+P^*S\rg^2-2|S|^2|P|^2\mbox{}
where
\bea  \label{As} A_1 &=&2 \ck e-\ck^4 -\ck^2(d-1)- b,~ A_2 = 1 - d
- \ck
(\ck + \ck c -2f),~ A_{12} = \ck \big(1 + d - \ck (\ck + f)+g\big)-e, \\
B_1 &=& 2 + \ck^4(2 + 4d - 3b) + b(4b-7) + 4\ck^5e -8\ck^3(2+
d)e -4\ck(2d + 4b-3)e + 4e^2  \nonumber \\
&& + 2\ck^2\big(5b +2d(d + 2b-1) + 6e^2-2\big), \label{B1}\\
B_2 &=& 2 + c + \ck^4(5c+2) + 4(d-1)d - 8\ck^3(c+1)f -
8\ck(c + 2d-1)f + 4f^2 + 2\ck^2\big((2c+1)(c + 2d-2) + 6f^2\big), \label{B2}\\
B_3 &=& 4\big[1 + \ck^6 + ( d-2)d + b(2d-1) + 4\ck^5f + 2e(e +
f)-2\ck\big(ce + 2(b-1)f + (2e + f)(2d + g)\big) + g^2 \nonumber\\
&&+\ck^4\big(c-2(d+g)-1\big)-2\ck^3\big(ce + f(4 + 2d +
g)\big)+\ck^2\big(b(1+2c)+4d+ c(2d-1)+ 2f(5e + f)\nonumber\\
&&+ 2g + (d + g)(3d+ g)-1\big)\big], \label{B3}\\
B_4 &=& 2\big[2\ck^4(f-3e) + 2e(2b + d + g-2) + \ck^5(2 + 4d +
g) +2\ck^3\big(b - d(4 + d + g) - 2(1 + e f + g)\big)\nonumber\\
&& +2\ck^2\big((2b + 2d-1)f + e(5 + 5d + 3g)\big) -
\ck\big(4e(2e+ f) - 3g + 2d(d + g-2 ) + 2b(2d + 2g+1 )-2\big)\big], \label{B4}\\
B_5 &=& 2\big[2\big((2d-1)e + f(d + g)\big)- 2\ck^4f -
2\ck^3\big(2f^2
+g +c(d + g+2)+2\big) + 2\ck^2\big(e + 2ce + f(2c + 5d + 3g+1)\big) \nonumber \\
&&+\ck^5(2 + 3c)+ \ck c(1 - 2d - 2g) + 2\ck\big(1 - 2f(2e + f)+ g
-2d(d +g)\big)\big], \label{B5}\eea and \bea B_6 &=& 2\ck^6 + 4e^2
+ 6\ck^5f + (4d-1)g - \ck^4(4 + 4d + 5g)
+4\ck^3\big(e - f(2 + d + g)\big) - 2\ck\big(f(2d + 2g-1)~~~~~~~~~~~~~~~~ \nonumber\\
&&+ e(2 + 4d + 4g)\big) +2\ck^2\big(2f(2e + f) + 3g + 2(d + d^2 +
 g d + g^2)+1\big).\label{B6}\eea

\end{widetext}

In this expansion, we neglect terms proportional to
$\sqrt{\Vhio\Vpqo}/\mP^2$, which are numerically suppressed. To
specify the canonically normalized scalar field, which could play
the role of inflaton, we have firstly to bring into a canonical
form, through a non-unitary transformation, the quadratic part
$K_{SP}$ of the K\"ahler potential in Eq.~(\ref{minK}) which
involves $S$ and $P$. Namely, we find
\beq K_{SP}=|S|^2+|P|^2+\ck (S P^*+ S^*P)= |\spl|^2+|\sm|^2,\eeq
where $s_{\pm}=A_{\pm}\lf S\pm P\rg/\sqrt{2}$ with
$A_{\pm}=\sqrt{1\pm\ck}$. Solving w.r.t. $S$ and $P$, we get
\beq\label{SPspm} S={1\over\sqrt{2}}\lf {s_+\over A_+}+{s_-\over
A_-}\rg~\mbox{and}~P={1\over\sqrt{2}}\lf {s_+\over A_+}-{s_-\over
A_-}\rg\cdot\eeq
Introducing the real and imaginary components of the fields
$s_{\pm}$ from the relations
$s_{\pm}=(s_{1\pm}+is_{2\pm})/\sqrt{2}$, the first term of $V_{\rm
HIs}$ in Eq.~(\ref{Vexpan}) reads~
\beq{\Vhio\over 2\mP^2}\Bigg(\llgm{s_{1-}&s_{1+}}\rrgm M^2_1
\matrix{\llgm{s_{1-} \cr s_{1+}}\rrgm} +\
(1\leftrightarrow2)\Bigg),\label{Vsugra} \eeq
where the matrices $M_1^2$ and $M_2^2$ are found to be
\beq M_1^2=M_2^2=\matrix{\llgm{ {A_1+A_2-2A_{12}\over
(1-\cks)^2(1+\cks)} & {A_1-A_2\over(1-\cks^2)^{3/2}}\cr
{A_1-A_2\over(1-\cks^2)^{3/2}} & {A_1+A_2+2A_{12}\over
(1+\cks)^2(1-\cks)}}\rrgm}\cdot\eeq
To identify the combination of $s_{\pm}$ which can play the role
of inflaton, we have to diagonalize the matrices $M_1^2$ and
$M_2^2$. This can be realized via an orthogonal matrix $U$ as
follows (\tr\ stands for the transpose of a matrix):
\beq U M_1^2 U^\tr =U M_2^2 U^\tr = \mbox{diag}\lf\msp,\msm \rg,
\eeq where \beq \label{mspm} \mspm={D_1 \pm \sqrt{D_2}\over(1 -
\ck^2)^2}\eeq with \beq D_1=A_1- 2\ck A_{12} + A_2\eeq and \beq
D_2=D_1^2 + 4(1-\ck^2)(A_{12}^2 - A_1A_2).\eeq Also \beq
U=\matrix{\llgm{ 1/\Np & 1/\Nm\cr \Vp/\Np & \Vm/\Nm}\rrgm}\eeq
with \beq U_{\pm}={\ck A_1 - 2A_{12} + \ck A_2 \pm
\sqrt{D_2}\over\sqrt{1 - \ck^2}(A_1 - A_2)}~\mbox{and}~
N_{\pm}=\sqrt{1+U^2_{\pm}}.\eeq
Embedding unity ($1=U U^\tr= U^\tr U$) on the left and the right
of $M^2_1$ and $M_2^2$ in Eq.~(\ref{Vsugra}), this expression can
be brought into the form
\beq
{\Vhio\over\mP^2}\lf\msp|\sg_{+}|^2+\msm\lfa\sg_-\rga^2\rg,\label{Vsugram}\eeq
where the real and imaginary components of the complex fields
$\sg_{+}$ and $\sg_{-}$, defined as $\sg_{\pm}
=(\sg_{1\pm}+i\sg_{2\pm})/\sqrt{2}$, are given by
\bea\sgam&=& s_{1-}/\Np+s_{1+}/\Nm ~\mbox{and}~\lf1\leftrightarrow2\rg,\label{sgm}\\
\sgap&=&s_{1-}\Vp/\Np+ s_{1+}\Vm/\Nm
~\mbox{and}~(1\leftrightarrow2).\label{sgp} \eea
Performing an appropriate R transformation, we can rotate $\sg_-$
to the real axis (setting $\sg_{2-}=0$). To simplify the notation,
we rename the remaining fields as follows: $\sgam=\sg,~~\sgap=s$
and $\sg_{2+}=q$. We then can solve Eqs.~(\ref{sgm}) and
(\ref{sgp}) w.r.t. the components of $s_\pm$, i.e.,
\bea && \label{s12m} s_{1-}=\Np\frac{s-\Vm\sg}{\Vp-\Vm}~\mbox{and}~ s_{2-}=\frac{\Np q}{\Vp-\Vm},~~~~\\
&& \label{s12p} s_{1+}=\Nm\frac{
s-\Vp\sg}{\Vm-\Vp}~\mbox{and}~s_{2+}=\frac{\Nm q}{\Vm-\Vp}\cdot
\eea
Substituting the expressions above into Eq.~(\ref{SPspm}) and then
into Eq.~(\ref{Vexpan}), we can derive $V_{\rm HIs}$ as a function
of $\sg, s,$ and $q$. As can be explicitly verified, $K_{SP}$
remains canonical w.r.t the latter fields -- i.e.
$K_{SP}=\lf\sg^2+s^2+q^2\rg/2$ -- since the $U$ transformation,
which connects $\sg, s,$ and $q$ with $s_{1\pm}$ and $s_{2\pm}$,
is orthogonal.

From the expressions above, we easily conclude that, if we set
$b=c=d=e=f=g=0$ with $\ck\neq0$, we take
\beq
\lf{A_1\over1-\ck^2},{A_{12}\over1-\ck^2},{A_2\over1-\ck^2}\rg=(\ck^2,\ck,1),
\eeq
which results to $\msm=0$ and $\msp=2$ via Eq.~(\ref{mspm}). In
other words, in this case, $\sg$ remains identically massless,
whereas $s$ and $q$ acquire effective masses equal to
$\sqrt{3/(1-\ck^2)}H_{\rm HI0}$ with $H_{\rm
HI0}=\sqrt{\Vhio}/\sqrt{3}m_{\rm P}$. Therefore, we are obliged to
invoke non-zero coefficients of the higher order terms of the \Ka\
considered in Eq.~(\ref{minK}), in order to generate a negative
$\msm$, and to reduce, thereby, $\ns$ to an observationally
acceptable level. Needless to say that, for $\ck=b=c=d=e=f=g=0$,
the well-known results (see e.g. Refs~\cite{effmass, moroiHm}) in
the context of minimal SUGRA can be obviously recovered -- note
that, in this case, $B_1=B_{2}=B_3/2=2$ and $B_4=B_5=B_6=0$.

\subsubsection{Radiative corrections}\label{fhi1c}

The constant tree-level potential energy density $\Vhio$ causes
SUSY breaking leading \cite{susyhybrid} to the generation of
one-loop radiative corrections, which provide a logarithmic slope
along the inflationary path necessary for driving the system
towards the vacua. Inserting the spectrum shown in
Table~\ref{tab1} in the well-known Coleman-Weinberg formula
\cite{cw}, we find (compare with Ref.~\cite{tetradis}) that the
one-loop radiative correction to $\Vhi$ is
\begin{equation} \label{Vcor} V_{\rm HIc}=V_{\kappa}+V_{\lambda},\eeq
where
\beq \label{Vk} V_\kappa = {\kappa^2 \Vhio\over
16\pi^2(1-\ck^2)}\left(2 \ln {\kappa^2x M^2
\over(1-\ck^2)\Lambda^2}+f_{\rm rc}(x)\right)\eeq
with $x=(1-\ck^2)|S|^2/M^2$ and \beq V_\lambda ={(\lambda-\ck\ka)
^2 \Vhio \over 32\pi^2(1-\ck^2)}\left(2 \ln {\kappa
(\lambda-\ck\ka) x_a M^2 \over(1-\ck^2)\Lambda^2}+f_{\rm
rc}\left(x_a\right) \right) \label{Vr} \eeq
%%%
with $x_a=(1-\ck^2)|\sigma_a|^2/\kappa(\lambda-\ck\ka) M^2$. In
the relations above, we have taken into account that the
dimensionality of the representations to which $\bar{\Phi}$ and
$\Phi$ [$\bar Q$ and $Q$] belong is 2~[1] -- see Table~I -- and we
have defined
\beq f_{\rm rc}(y)=(y+1)^{2}\ln\left(1+{1/y}\right)
+(y-1)^{2}\ln\left(1-{1/y}\right).\label{frc} \eeq
Let us note that, for $x\gg1$ and $x_a\gg1$, $V_\kappa$
[$V_\lambda $] can be well approximated by replacing $f_{\rm
rc}(y)\simeq3$ in the RHS of Eq.~(\ref{Vk}) [Eq.~(\ref{Vr})] --
see Sec.~\ref{fhi1dyn}. Although rather large $\kappa$'s,
$\kappa_a$'s, and $\lambda$'s are used in our work,
renormalization group effects \cite{espinoza} remain negligible
and so our results are independent from the renormalization scale
$\Lambda$.

Substituting Eqs.~(\ref{s12m}) and (\ref{s12p}) into
Eq.~(\ref{SPspm}) and then into Eqs.~(\ref{Vk}) and (\ref{Vr}), we
can derive $V_{\rm HIc}$ as a function of $\sg, s,$ and $q$.

\subsection{The inflationary dynamics}\label{fhi1dyn}

The evolution of the various fields involved in our scheme during
FHI is governed by their {\it equations of motion} (e.o.m.)
\beq\ddot f+3H\dot f+V_{{\rm
HI},f}=0~\mbox{with}~f=\sigma,~s,~\mbox{and}~q. \label{eqf}\eeq
where the dot denotes derivation w.r.t. the cosmic time $t$ and
$H$ is the Hubble parameter. The solution of the system in
Eq.~(\ref{eqf}) can be facilitated if we use as independent
variable the number of e-foldings $N$ defined by
\beq N=\ln \left(R/R_{\rm HIi}\right)~\Rightarrow~\dot N
=H~\mbox{and}~\dot H=H'H\label{Ndfn} \eeq
with $R(t)$ being the scale factor of the universe and $R_{\rm
HIi}$ its value at the commencement of FHI. Here the prime denotes
derivation w.r.t. $N$. Converting the time derivatives to
derivatives w.r.t. $N$, Eq.~(\ref{eqf}) becomes
\beq H^2f''+(3H+H')Hf'+V_{{\rm HI},f}=0 \label{eqfN}\eeq
with $f=\sigma,~s,$ or $q$. This system can be solved numerically
for the period of FHI by taking
\beq H=H_{\rm HI}=\sqrt{2\Vhio\over{6m^2_{\rm
P}-\sigma'^2-s'^2-q'^2}},\eeq
$V=\Vhi$ given by Eq.~(\ref{Vol}), and imposing the following
initial conditions (at $N=0$):
\beq f(0)=f_{\rm HIi}~\mbox{and}~
f'(0)=0~\mbox{with}~f=\sigma,~s,~\mbox{or}~q, \label{eqfin}\eeq
where $f_{\rm HIi}$ is taken to be in the range
$(1.5-4.5)\times10^{17}~\GeV$. We checked that our results are
pretty stable against variation of $f_{\rm HIi}$ -- see also
Sec.~\ref{num}.

Nevertheless, we can obtain a comprehensive and rather accurate
approximation to the inflationary dynamics if we put $H\simeq
H_{\rm HI0}=\sqrt{\Vhio}/\sqrt{3}\mP$ ($H'=0$) and keep in
Eq.~(\ref{eqfN}) the most important terms in the expansion of
$V_{{\rm HI},f}$. In particular, we can easily verify that
$V_{{\rm HI},q}$ turns out to be proportional to $q$. As a
consequence, Eq.~(\ref{eqfN}) for $f=q$ is a second order linear
homogeneous differential equation which admits an oscillatory
solution with decreasing amplitude. Therefore, $q$ rapidly
decreases to zero and so it does not influence the dynamics of the
other fields. On the contrary, the e.o.m. of $s$ is non-homogenous
as we can realize by inserting into Eq.~(\ref{eqfN}) for $f=s$ the
expression
\beq \label{Vs} V_{{\rm HI},s}\simeq \Vhio\left({\msp\over\mP^2}
s-{2\kp^2C_1+(\ld-\ck\ka)^2C_2\over8(1-\ck^2)\pi^2\sg}\right),\eeq
where the last two terms in the RHS of this equation have been
derived by expanding the exact result for $s/\sigma\rightarrow 0$.
Here
\beq C_1=\frac{N_S}{D_S}~\mbox{and}~C_2=\frac{\ld^2 N_S+\ka^2
N_P+2\ld\ka N_{SP}}{\ld^2 D_S+\ka^2 D_P+2\ld\ka D_{SP}},\eeq
where we have used the abbreviations:
\bea && \nonumber N_{S[P]} = {\Np^2 \Vm\over\Am^2} + {\Nm^2
\Vp\over\Ap^2}-[+]{\Nm \Np (\Vm + \Vp)\over\Ap\Am},\\
&&  \nonumber N_{SP} = {\Nm^2 \Vp\over\Ap^2}-{\Np^2
\Vm\over\Am^2},~
D_{SP} = {\Nm^2 \Vp^2\over\Ap^2}-{\Np^2 \Vm^2\over\Am^2},\\
&&  D_{S[P]} = \lf {\Np \Vm\over \Am} -[+]~{\Nm \Vp\over\Ap}\rg^2.
~~~~~~~~~~~~~~~~ \eea
Since the general solution of the corresponding homogenous
differential equation rapidly decreases to zero, as the field $q$,
the solution of Eq.~(\ref{eqfN}) with $f=s$, is dominated by the
following particular solution
\beq s\simeq
{2\kp^2C_1+(\ld-\ck\ka)^2C_2\over8(1-\ck^2)\pi^2\msp\sg}\mP^2,\label{ssol}\eeq
which minimizes $\Vhi$ in the $s$-direction, as can be seen from
Eq.~(\ref{Vs}). In sharp contrast to the situation of
Refs.~\cite{kawasaki, newkawasaki, yamaguchi}, in our case $s$
turns out to be just mildly, and not drastically, reduced w.r.t.
$\sigma$.

In the slow-roll approximation, which is determined by the
condition
\beq \label{slow} {\sf
max}\{\epsilon(\sigma),|\eta(\sigma)|\}\leq1,\nonumber\eeq
where\beq\epsilon\simeq{m^2_{\rm P}\over2}\left(\frac{V_{\rm
HI,\sigma}}{V_{\rm HI}}\right)^2~\mbox{and}~\eta\simeq m^2_{\rm
P}~\frac{V_{\rm HI,\sigma\sigma}}{V_{\rm HI}},\eeq
the e.o.m. of $\sigma$ in Eq.~(\ref{eqfN}) takes the form
$-3H^2\sigma'\simeq V_{\rm HI,\sg}$ or
\beq -\sg'\simeq\msm\sg+{\mP^2(2\kp^2+
(\ld-\ck\ka)^2)\over8\pi^2(1-\ck^2)\sg}+{C_3\sg^3\over\mP^2(1-\ck^2)^2},
\label{eomsg}\eeq
where -- keeping only the most important terms in the expansion of
the second term of the RHS of Eq.~(\ref{Vexpan}) -- we have
\bea \nonumber C_3&=&{1\over16(\Vm-\Vp)^4}\big( B_1 D_S^2+B_2 D_P^2\\
&&  +\ B_3 D_{SP}^2+2B_4 D_{SP} D_S\big).\label{C3}\eea
Solving Eq.~(\ref{eomsg}) w.r.t. $\sigma$, we get
\bea \nonumber \sg&= &{\mP(1 - \ck^2)\over\sqrt{2 C_3}}
\Bigg[-\msm - D\tan\bigg(D N \\ && - \arctan {1\over D}\lf\msm +
{2 C_3 \sigma_{\rm HIi}^2\over \mP^2(1 -
\ck^2)^2}\rg\bigg)\Bigg]^{1/2} \label{sgsol}\eea
with $\sigma_{\rm HIi}$ being the initial value of $\sigma$ at the
beginning of FHI and
\beq D = \sqrt{C_3\big(2\kp^2 + (\ld -\ck \ka)^2\big)/2\pi^2(1 -
\ck^2)^3-m_-^4 }.\eeq
The end of FHI takes place at $\sigma=\sigma_{\rm f}$, when the
conditions of Eq.~(\ref{flat}) or Eq.~(\ref{slow}) are violated.
For $\kp\lesssim0.005$, one can consider the former case only
since the slow-roll conditions are violated infinitesimally close
to the relevant critical point of the inflationary trajectory.

Soon afterwards, the inflaton system consisting of the two complex
scalar fields $S$ and $(\delta\bar \Phi+\delta \Phi)/\sqrt{2}$
(where $\delta\bar \Phi=\bar \Phi-M$ and $\delta\Phi=\Phi-M$) with
mass $m_{1\rm inf}=\sqrt{2}\kappa M$ settles into a phase of
damped oscillations about the SUSY vacuum and decays to MSSM
degrees of freedom reheating the universe. The predominant decay
channels of $S$ and $(\delta\bar \Phi+\delta \Phi)/\sqrt{2}$ are
to fermionic and bosonic $\bar Q,~Q$ respectively via tree-level
couplings derived from the last term in the RHS of Eq.~(\ref{Whi})
with a common decay width
\beq \Gamma_1={1\over16\pi}\lambda ^2\,m_{\rm inf} ^2.
\label{gammas}\eeq
The corresponding reheat temperature is \cite{quin} given by
\beq T_{\rm 1rh}=\left(72\over5\pi^2g_{\rm 1rh*}\right)^{1/4}
\sqrt{\Gamma_1 m_{\rm P}},\label{T1rh}\eeq
where $g_{\rm 1rh*}=g_{*}(T_{\rm 1rh})$ counts the effective
number of relativistic degrees of freedom at temperature $T_{\rm
1rh}$. We find $g_{\rm 1rh*}\simeq438.75$~[$g_{\rm
1rh*}\simeq513.75$] for the MSSM spectrum plus right handed
neutrinos and the particle content of the superfields $P$, $\bar
Q$, $Q$, $\bar D_{\rm a}$, $D_{\rm a}$, and $H_{\rm a}$ for $n=5$
[$n=7$].

For relatively large $\ld$'s $(\simeq0.05-0.1)$, we get $T_{\rm
1rh}>V_{\rm PQ0}^{1/4}$. As a consequence, after the end of FHI,
we obtain \emph{matter domination} (MD) for $T\geq T_{\rm 1rh}$
and \emph{radiation domination} (RD) for $\Vpqo^{1/4}\lesssim
T\lesssim T_{\rm 1rh}$. During MD the evolution of $s$ can be
found by solving its e.o.m. for $N>N_{\rm HI}$, i.e., inserting
\cite{newkawasaki, yamaguchi, ltetradis}
\beq H=H_{\rm HI0}e^{-3(N-N_{\rm HI})/2},~V={3H^2\over4}s^2
\label{MDs} \eeq
into Eq.~(\ref{eqfN}). Taking also into account that, during the
MD epoch, $R\propto\rho_{\rm osc}^{-1/3}$ (where $\rho_{\rm osc}$
is the energy density of the oscillating system), we can derive
the value of $s$ at the beginning of PQPT:
\beq s_{\rm PQi}\simeq\left({\rho_{\rm 1rh}\over
\Vhio}\right)^{1/4}s_{\rm HIf}~\mbox{with}~\rho_{\rm
1rh}={\pi^2\over30}g_{\rm 1rh*}T_{\rm 1rh}^4\label{s2i}\eeq
being the radiation energy density at temperature $T_{\rm 1rh}$
and $s_{\rm HIf}$ the value of $s$ at the end of FHI, which can be
approximated by inserting $\sigma=\sigma_{\rm f}$ into
Eq.~(\ref{ssol}). Note that during the subsequent RD era, $s$
remains frozen (see Ref.~\cite{moroiHm} and footnote 1 in
Ref.~\cite{qlocked}) and so the further reduction of $s$ relative
to $\sg$ can be evaded -- cf. Ref.~\cite{kawasaki}.

\section{The stage of PQPT}\label{fhi2}

For $T\lesssim\Vhio^{1/4}$, the cosmological dynamics is governed
by the second term of $W$ in the RHS of Eq.~(\ref{Whi}). The SUGRA
corrections can be safely ignored since $|P|\sim f_a\ll \mP$. The
relevant F-term scalar potential is
\beq\label{Vpqf} V_{\rm PQF}=\kappa_a^2\left|\bar Q
Q-f^2_a/4\right|^2\ +\ \kappa^2_a\left|P\right|^2\left(|\bar Q|^2+
|Q|^2\right).\eeq
Along the F--flat direction in Eq.~(\ref{PQflat}), $V_{\rm PQF}$
takes the constant value in Eq.~(\ref{V2}), which breaks SUSY
creating a mass splitting in the supermultiples $Q$ and $\bar Q$
and giving rise to one-loop radiative correction, $V_{\rm PQc}$,
to the relevant potential -- cf. Sec.~\ref{fhi1c}. Indeed, the
particle spectrum there includes a Dirac spinor with mass $\ka
|P|$ and 2 complex scalars with mass squared $\ka^2\lf|P|^2\pm
f_a^2/4\rg$ and, therefore, $V_{\rm PQc}$ can be written \cite{cw}
as follows:
\beq \label{Vpqrc} V_{\rm PQc}= {\kappa_a^2 \Vpqo\over
32\pi^2}\left(2 \ln {\kappa_a^2A_P^2s^2 \over\Lambda^2}+f_{\rm
rc}(x_s)\right),\eeq where \beq A_P={\Am\ \Nm+\Ap\ \Np\over\Ap\Am
(\Vm-\Vp)}\eeq
with $x_s=A_P s/f_a$ and $P=A_P s/2$ -- see Eqs.~(\ref{SPspm}),
(\ref{s12m}), and (\ref{s12p}). When $|P| < f_a/2$, one mass
squared becomes negative and suggests a phase transition along the
$|P|$-axis. Assuming gravity mediated soft SUSY breaking, the
potential there has the form
\beq\label{Vpq} V_{\rm PQ}=\Vpqo+{1\over2}m^2_s\,s^2-
\sqrt{2\Vpqo}\;|{\rm a}_s| s+V_{\rm PQc}\eeq
for $s\geq f_a/A_P$. Here, $m_s$ is the soft SUSY breaking mass of
$s$ and ${\rm a}_s$ is the soft SUSY breaking tadpole
\cite{sstad}. In Eq.~(\ref{Vpq}), we take the phase of ${\rm a}_s$
to be $\arg({\rm a}_s)=\pi$ which minimizes the $V_{\rm PQ}$ for
given $s>0$.  For reasonable values of the parameters involved,
$V_{\rm PQ}$ does not give rise to inflation, mainly due to the
contribution of $V_{\rm PQc}$ which spoils the $\eta$-criterion --
and has not been taken into account in the analysis of
Ref.~\cite{hybrid}. This negative result remains even if we
consider dissipative effects \cite{berera} due to the presence of
the extra superpotential terms in Eq.~(\ref{Wdw}).

Therefore, we are obliged to assume that after a negligible number
of e-folds, the system consisting of the two complex scalar fields
$P$ and $(\delta\bar Q+\delta Q)/\sqrt{2}$ (where $\delta\bar
Q=\bar Q-f_a/2$ and $\delta Q=Q-f_a/2$) with mass $m_{\rm
PQ}=\kappa_a f_a/\sqrt{2}$ enters into an oscillatory phase about
the PQ minimum and eventually decays, via the first
non-renormalizable coupling in the RHS of Eq.~(\ref{Wmu}), to
Higgses and Higgsinos respectively with a common decay width
\cite{lectures}
\beq \Gamma_{2}={1\over2\pi}\lambda_\mu^2\left({f_a\over 2m_{\rm
P}}\right)^2m_{\rm PQ}. \label{2gammas}\eeq
The corresponding reheat temperature is \cite{quin} calculated by
\beq T_{\rm 2rh}=\left(72\over5\pi^2g_{\rm 2rh*}\right)^{1/4}
\sqrt{\Gamma_{2} m_{\rm P}},\label{T2rh}\eeq
where $g_{\rm 2rh*}=228.75$ is the energy density effective number
of degrees of freedom for the MSSM spectrum.

\section{Observational constraints}\label{cont}

Under the assumptions that (i) the curvature perturbation
generated by $\sigma$ is solely responsible for the observed
curvature perturbation and (ii) the violation of Eq.~(\ref{flat})
occurs along the lines of Sec.~\ref{inflation12}, the parameters
of our model can be restricted imposing the following
requirements:

\paragraph{} According to the inflationary paradigm, the horizon and flatness
problems of the \emph{standard Big Bang} (SBB) can be successfully
resolved provided that the number of e-foldings, $N_{\rm HI*}$,
that the scale $k_*=0.002/{\rm Mpc}$ suffers during FHI takes a
certain value which depends on the details of the cosmological
scenario. Employing standard methods \cite{hinova}, we can easily
derive the required $N_{\rm HI*}$ at $k_*$:
\beq N_{\rm HI*}\simeq23+{2\over 3}\ln{V^{1/4}_{\rm
HI0}\over{1~{\rm GeV}}}-{1\over 3}\ln{V^{1/4}_{\rm
PQ0}\over{1~{\rm GeV}}}+ {1\over3}\ln {T_{\rm 1rh}T_{\rm
2rh}\over{1~{\rm GeV}^2}}\label{Ntott} \eeq
consistently with the fact that $T_{1\rm rh}>V_{\rm PQ0}^{1/4}$
and, thus, we obtain MD followed by RD during the era between the
end of FHI and the onset of PQPT. On the other hand, $N_{\rm HI*}$
can be found from
\beq N_{\rm HI*}=N_{\rm HI}-N_*,\label{N1}\eeq
where $N_{\rm HI}$ is the total number of e-foldings generated
during FHI and $N_*$ is the number of e-foldings elapsed from the
onset of FHI until the scale $k_*$ crosses outside the horizon of
FHI and corresponds to the field value $\sigma_{*}$.

\paragraph{} The power spectrum $P_{\cal R}$ of the curvature
perturbation, which is generated during FHI and can be calculated
at the pivot scale $k_{*}$ as a function of $\sg_*$, is to be
confronted with the WMAP7 data \cite{wmap3}, i.e.
\begin{equation} \label{Prob}
P^{1/2}_{\cal R}= \frac{1}{2\sqrt{3}\, \pi m^3_{\rm P}}\;
\left.\frac{V_{\rm HI}^{3/2}}{|V_{{\rm HI},\sigma}
|}\right\vert_{\sigma=\sigma_*}\simeq\: 4.93\times 10^{-5}.
\end{equation}

\paragraph{}  The (scalar) spectral index $n_{\rm s}$, which is
given by
\beq \label{nS} n_{\rm s}=1-6\epsilon(\sigma_*)\ +\
2\eta(\sigma_*),\eeq
is to be consisted with the fitting of the WMAP7 data by the
standard power-low $\Lambda$CDM:
\begin{equation}\label{nswmap}
n_{\rm s}=0.963\pm0.028~\Rightarrow~0.935\lesssim n_{\rm s}
\lesssim 0.991
\end{equation}
at 95$\%$ \emph{confidence level} (c.l.). Note, in passing, that
the running of $\ns$ and the scalar-to-tensor ratio turn out to be
vanishingly small in our model as in any model of FHI.

\paragraph{} In order for the PQPT to take place after a short
temporary domination of $\Vpqo$, the value of $|P|$ at the onset
of PQPT, $\left|P_{\rm PQi}\right|$, must be adequately larger
than its critical value along the F-flat direction in
Eq.~(\ref{PQflat}). This entails
\beq \left|P_{\rm PQi}\right|>f_a/2~\Rightarrow~s_{\rm
PQi}>f_a/A_P,\label{s2ic}\eeq
which implies mainly a lower bound on the parameter $\lambda$
through Eqs.~(\ref{T1rh}) and (\ref{s2i}). As we emphasize in
Sec.~\ref{model}, this requirement is directly related to the
reduction of the $\Gr$ yield at the onset of nucleosynthesis,
$Y_{\Gr}$, to an acceptable level. Indeed, $Y_{\Gr}$ in the case
that $\Vpqo$ does not dominate can be estimated \cite{kohri} as
\beq\label{Y1}
Y_{1\Gr}\simeq1.9\times10^{-12}\left(\Trha/10^{10}~{\rm
GeV}\right).\eeq
However, in the opposite case, if we take into account the entropy
produced after PQPT during the subsequent reheating process -- for
computational details see Appendix~\ref{Rhg} -- we obtain
\beq\label{Y2} Y_{2\Gr}\simeq\lf{\pi^2\over 30}g_{\rm
1rh*}\rg^{1/4}{\Trhb\over\Vpqo^{1/4}}Y_{1\Gr},\eeq
which is suppressed relative to $Y_{1\Gr}$. In order to avoid
spoiling the success of the SBB nucleosynthesis, an upper bound on
$Y_{\Gr}$ is to be imposed depending on the $\Gr$ mass, $m_{\Gr}$,
and the dominant $\Gr$ decay mode. For the conservative case that
$\Gr$ decays with a tiny hadronic branching ratio, we have
\beq\label{Ymax} Y_{\Gr}\lesssim\left\{\matrix{
%\begin{array}{rl}
10^{-14}\hfill \cr
10^{-13}\hfill \cr
10^{-12}\hfill \cr}
%\end{array}
\right.~\mbox{for}~m_{\Gr}\simeq\left\{\matrix{
%\begin{array}{rl}
0.69~{\rm TeV}\hfill \cr
10.6~{\rm TeV}\hfill \cr
13.5~{\rm TeV}\hfill \cr}
%\end{array}
\right.\eeq
respectively. The bound above can be somehow relaxed in the case
of a stable $\Gr$. However, it is achievable in our model, as we
see below.

\section{Numerical results}\label{num}

As can be easily seen from the relevant expressions above, our
cosmological set-up depends on the following parameters:
$$\kappa,~\ka,~\ld,~M,~f_a,~\lambda_\mu,~n,~\ck,~b,~c,~d,~e,~f,~\mbox{and}~g.$$
We fix $f_a=10^{12}~\GeV$ and $\lambda_\mu=0.01$ so as to obtain
$\mu\sim1~{\rm TeV}$. We also set $n=5$. These three parameters
are involved in the determination of $T_{\rm 1rh}$ and $T_{\rm
2rh}$ -- via Eq.~(\ref{T1rh}) and (\ref{T2rh}) respectively -- and
play no crucial role in the inflationary dynamics. We also choose
$\ka=0.01$ and $\ld=0.1$. Variation of $\ka$ leads to a variation
of $s$ according to Eq.~(\ref{ssol}) without creating drastic
changes in the inflationary predictions for $M,~N_{\rm HI*},$ and
$\ns$. On the other hand, $\ld$ controls crucially $s$ and $T_{\rm
1rh}$ through Eqs.~(\ref{ssol}) and (\ref{s2i}) and the upper
bound on the condition of Eq.~(\ref{flat}{b}). Finally, we fix
throughout our numerical computation $c=d=e=f=g=0.1$ since these
parameters -- contrary to $\ck$ and $b$ -- have a minor impact on
the calculation of $\msp$ and $\msm$. As we show below, the
selected values above give us a wide and natural allowed region of
the remaining fundamental inflationary parameters ($\kp, M, \ck,$
and $b$). Besides the parameters above, in our computation, we use
as input parameters the quantities $N_*$ and $f_{\rm HIi}=I$ with
$f=\sg,~s,$ and $q$. We set $I\simeq(1.5-4.5)\times10^{17}~\GeV$
so as to obtain $N_{\rm HI}\simeq70-100$. We then restrict $M$ and
$N_*$ so that Eqs.~(\ref{Ntott}) and (\ref{Prob}) are fulfilled.
We check also if the violation of Eqs.~(\ref{flat}{a}) and
Eq.~(\ref{flat}{b}) occurs according to the desired order so that
our scenario is realized successfully. Using Eqs.~(\ref{s2i}) and
(\ref{nS}), we can extract $s_{\rm PQi}$ and $n_{\rm s}$ and
compare them with the requirements of Eqs.~(\ref{s2ic}) and
(\ref{nswmap}). For the solutions presented below,
Eq.~(\ref{s2ic}) is safely fulfilled.

\begin{figure}[t]
\includegraphics[width=60mm,angle=-90]{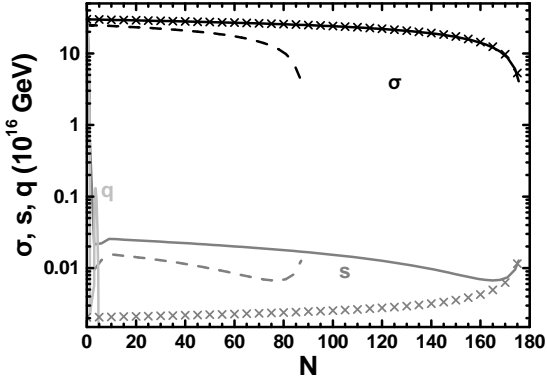}
\caption{The evolution of $\sg$ (black lines), $s$ (gray lines),
and $q$ (light gray lines) as evaluated via the solution of
Eq.~(\ref{eqfN}) as functions of $N$ for
$\kp=0.002,~\ka=-b=0.01,~\ck=-0.0125,~\ld=c=d=e=f=g=0.1,$ and
$\sg_{\rm HIi}=s_{\rm HIi}=q_{\rm HIi}=3\times10^{17}~\GeV$ (solid
lines) or $\sg_{\rm HIi}=s_{\rm HIi}=q_{\rm
HIi}=2.5\times10^{17}~\GeV$ (dashed lines). Crosses are obtained
by applying our analytical approach for $\sg_{\rm HIi}=s_{\rm
HIi}=q_{\rm HIi}=3\times10^{17}\GeV$. }\label{fN}
\end{figure}

The selected $I$ is to be large enough so that $s$ reaches the
attractor of Eq.~(\ref{ssol}). Under this assumption, our results
are independent of the precise $I$, as can be clearly deduced from
Fig.~\ref{fN}, where we plot $\sg$ (black lines and crosses), $s$
(gray lines and crosses), and $q$ (light gray lines) as functions
of $N$ for $\kp=0.002,~\ck=-0.0125,~b=-0.01,$ and $\sg_{\rm
HIi}=s_{\rm HIi}=q_{\rm HIi}=2.5\times10^{17}~\GeV$ (dashed lines)
or $\sg_{\rm HIi}=s_{\rm HIi}=q_{\rm HIi}=3\times10^{17}~\GeV$
(solid lines and crosses). The lines are drawn by solving
numerically Eq.~(\ref{eqfN}), whereas crosses are obtained by
employing Eqs.~(\ref{ssol}) and (\ref{sgsol}). For both choices of
$\sg_{\rm HIi}=s_{\rm HIi}=q_{\rm HIi}$'s, we obtain
$m_-^2=-0.0136$, $m_+^2=1.82$, $N_{\rm HI*}=52.5$,
$M=2.86\times10^{16}~\GeV$, $\ns=0.963$, $\sg_{\rm
HIf}=4.08\times10^{16}~\GeV$, and $s_{\rm
HIf}=1.3\times10^{14}~\GeV$ although in the first [second] case we
obtain $N_{\rm f}=87.1$ [$N_{\rm f}=175.5$], where $N_{\rm f}$ is
the number of e-foldings elapsed from the commencement of FHI
until Eq.~(\ref{flat}{a}) is violated. It is impressive that $M$
can take a value exactly equal to the SUSY GUT breaking scale,
$M_{\rm GUT}=2.86\times10^{16}~{\rm GeV}$, contrary to all other
realizations of the standard FHI -- cf. Refs.~\cite{susyhybrid,
mhi, king, tetradis}.  Despite the fact that
$T_{\rm1rh}=4.7\times10^{13}~\GeV$, $s_{\rm PQi}/A_Pf_a\simeq23$
and, therefore, a second reheating process is possible which
results to $T_{\rm2rh}=3\times10^{4}~\GeV$. It is worth
emphasizing that $Y_{1\Gr}\simeq9\times 10^{-9}$, whereas
$Y_{2\Gr}\simeq1.9\times 10^{-14}$ which is consistent with the
constraint of Eq.~(\ref{Ymax}). If we set $\lambda_\mu=0.05$, we
take $\mu\simeq5~{\rm TeV}$, $T_{\rm2rh}=1.5\times10^{5}~\GeV,$
and $Y_{2\Gr}\simeq9.5\times 10^{-14}$ which again falls into the
ranges of Eq.~(\ref{Ymax}). Finally, we observe that our
analytical findings on $\sg$ are very close to the numerical ones
for all $N$'s. On the contrary, analytical and numerical results
on $s$ converge mainly for $N\simeq N_{\rm f}$. Let us clarify
that the results presented in the following are derived
exclusively by our numerical program.

\begin{figure}[!t]
\includegraphics[width=60mm,angle=-90]{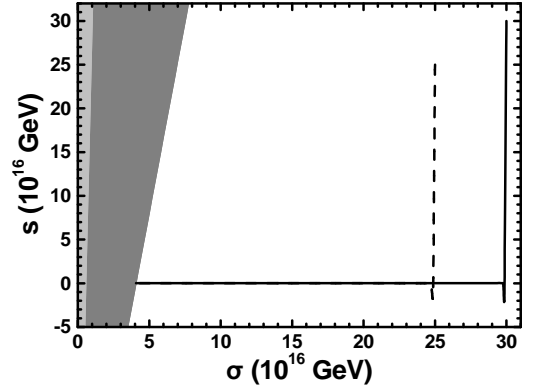}
\caption{The evolution of $\sg$ and $s$ as evaluated via the
solution of Eq.~(\ref{eqfN}) in the $\sg-s$ plane for
$\kp=0.002,~\ka=-b=0.01,~\ck=-0.0125,~\ld=c=d=e=f=g=0.1,$ and
$\sg_{\rm HIi}=s_{\rm HIi}=3\times10^{17}~\GeV$ (solid line) or
$\sg_{\rm HIi}=s_{\rm HIi}=2.5\times10^{17}~\GeV$ (dashed line).
The gray [light gray] region is excluded by Eq.~(\ref{flat}{a})
[Eq.~(\ref{flat}{b})].}\label{sgs}
\end{figure}

\begin{figure*}[!t]
\centering
\includegraphics[width=60mm,angle=-90]{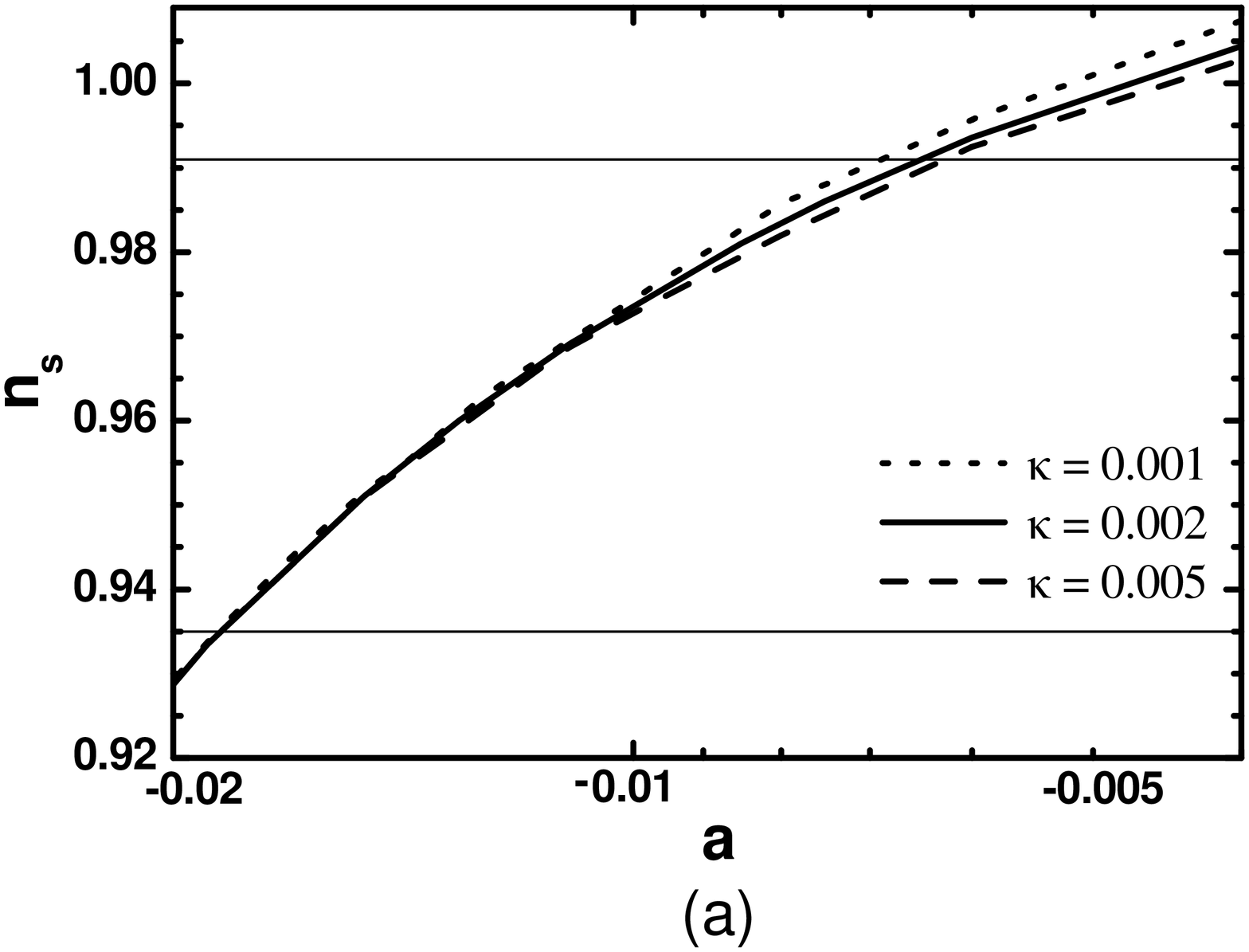}
%\hspace{-.1mm}
\includegraphics[width=60mm,angle=-90]{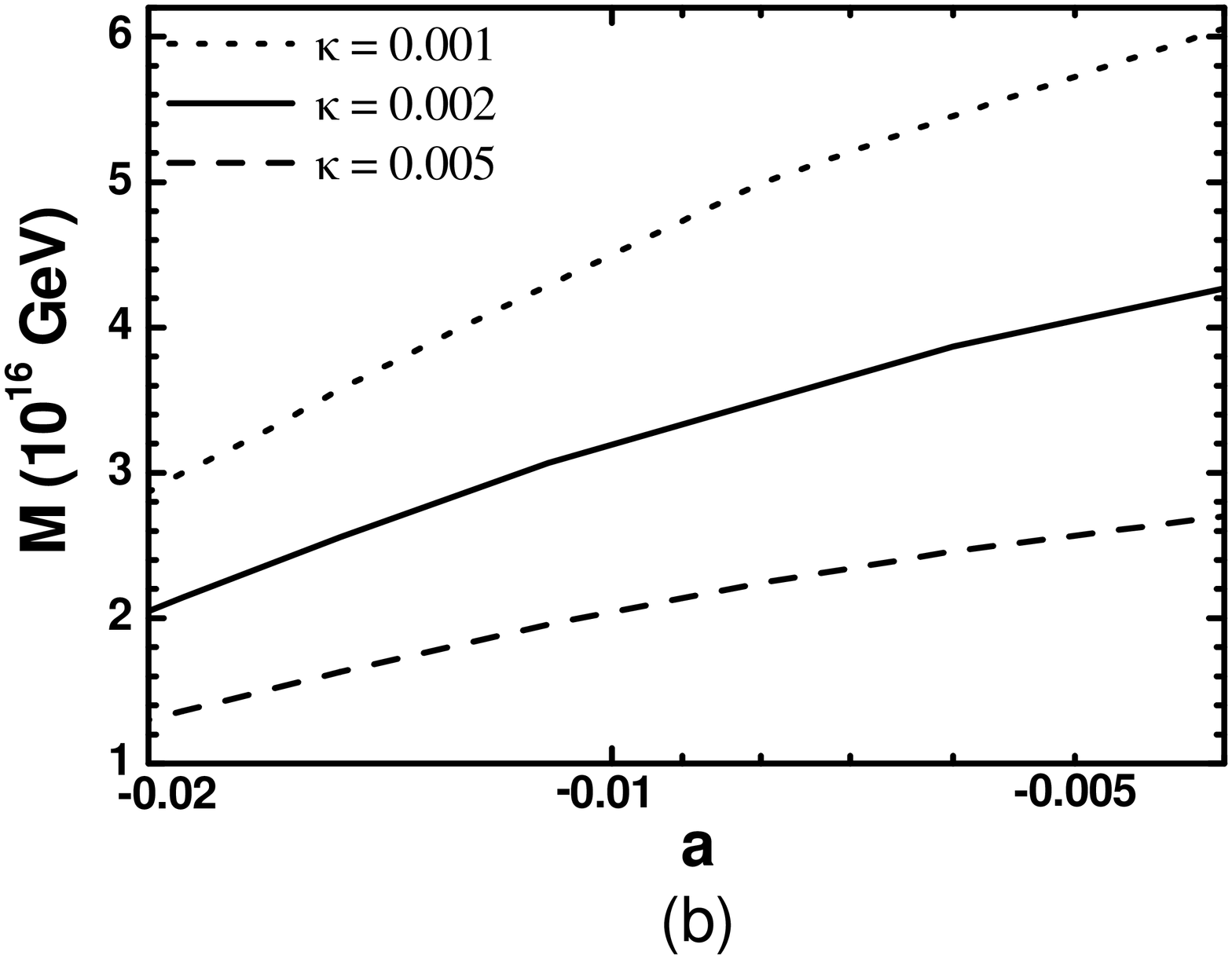}
\caption{\label{anM} The allowed by Eqs.~(\ref{Ntott}) and
(\ref{Prob}) values of $\ns$ {(a)} and $M$ {(b)} versus $\ck$ for
$\ka=-b=0.01,~\ld=c=d=e=f=g=0.1,$ and several $\kappa$'s indicated
in the graphs. The region of Eq.~(\ref{nswmap}) is also limited by
thin lines in (a).}
\end{figure*}

As observed in Fig.~\ref{fN}, immediately after the onset of FHI,
$q$ decreases sharply and so it does not influence the
inflationary dynamics. The evolution of the other two fields
($\sg$ and $s$) can be represented in the $\sg-s$ plane as in
Fig.~\ref{sgs}, where we show the evolution of $\sg$ and $s$ for
the parameters used in Fig.~\ref{fN} and taking $\sg_{\rm
HIi}=s_{\rm HIi}=3\times10^{17}~\GeV$ (solid line) or $\sg_{\rm
HIi}=s_{\rm HIi}=2.5\times10^{17}~\GeV$ (dashed line). There, we
draw also the gray [light gray] region which is excluded by
Eq.~(\ref{flat}{a}) [Eq.~(\ref{flat}{b})]. We remark that
Eq.~(\ref{flat}{a}) is violated earlier and so the preferred
hierarchy in the domination of $\Vhio$ or $\Vpqo$ in our
cosmological proposal is valid -- see Sec.~\ref{model}.

The importance of the coefficient $\ck$ in reducing $\ns$ can be
easily concluded from Fig.~\ref{anM}(a), where we depict $\ns$
versus $\ck$ for $b=-0.01$ and various $\kp$'s indicated in the
graph. Increasing the absolute value of $\ck$, $|\msm|$ increases
too and so $\ns$ decreases and becomes consistent with the
observationally favored range of Eq.~(\ref{nswmap}). On the other
hand, the same variation of $\ck$ leads to a reduction of $M$
which lies around its SUSY GUT value, $M_{\rm GUT}$, as can be
concluded from Fig.~\ref{anM}(b).

\begin{figure*}[!tb]
\centering
\includegraphics[width=60mm,angle=-90]{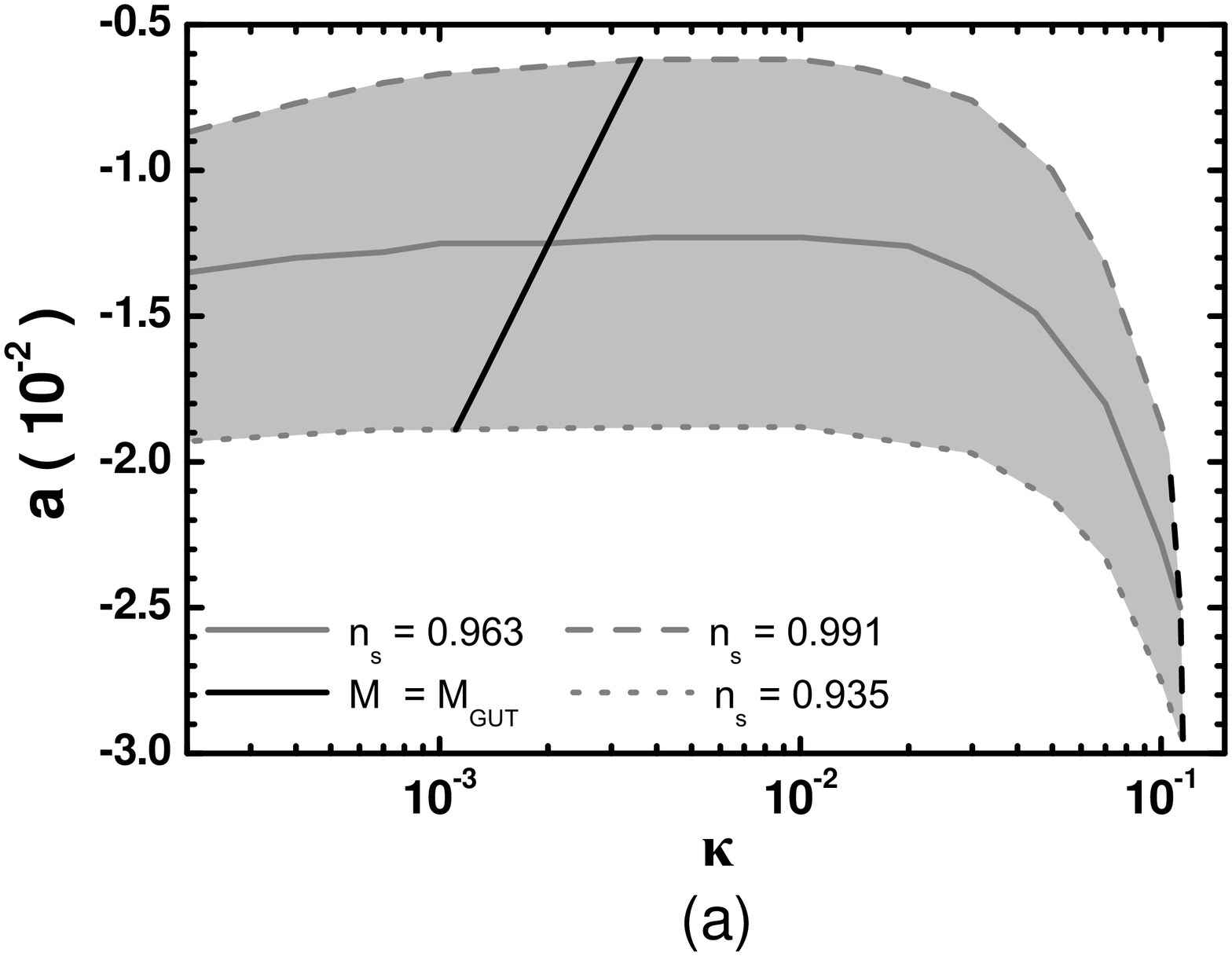}
%\hspace{-.1mm}
\includegraphics[width=60mm,angle=-90]{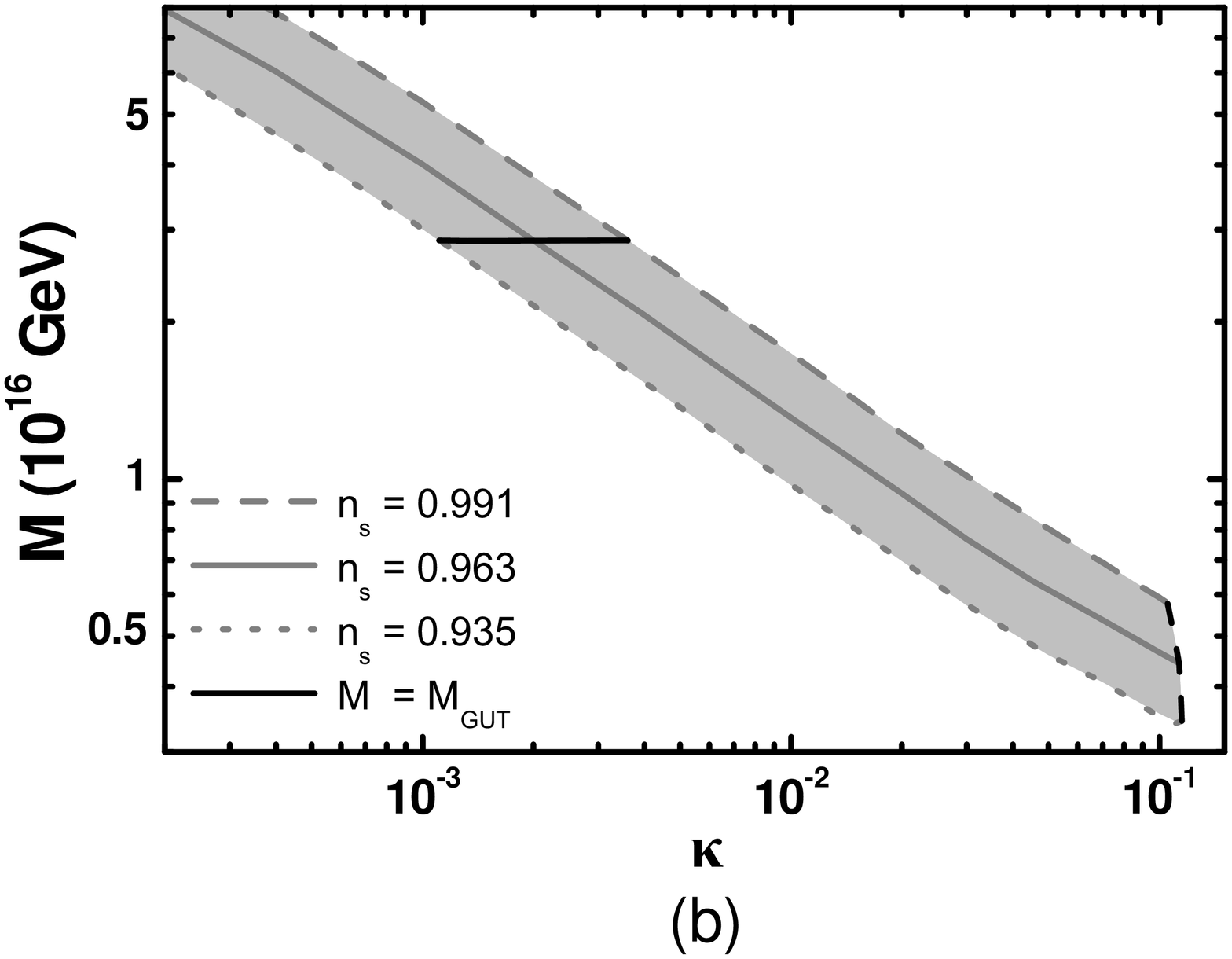}
\caption{\label{kaM} Allowed (lightly gray shaded) regions in the
{(a)} $\kappa-\ck$ and {(b)} $\kappa-M$ plane for $\ka=-b=0.01$
and $\ld=c=d=e=f=g=0.1$. The gray dashed [dotted] lines correspond
to the upper [lower] bound on $n_{\rm s}$ in Eq.~(\ref{nswmap}),
whereas the gray solid lines have been obtained by fixing $n_{\rm
s}$ to its central value in Eq.~(\ref{nswmap}). Along the black
solid lines, we fix $M=M_{\rm GUT}$, whereas beyond the black
dashed lines the preferred hierarchy in the violation of
Eqs.~(\ref{flat}{a}) and (\ref{flat}{b}) is broken. In the allowed
regions, Eqs.~(\ref{Ntott}) and (\ref{Prob}) are also fulfilled. }
\end{figure*}

Confronting FHI with all the constraints of Sec.~\ref{cont}, we
can delineate the allowed (lightly gray shaded) region in the
$\kappa-\ck$ [$\kappa-M$] plane as in Fig.~\ref{kaM}(a)
[Fig.~\ref{kaM}(b)], where we take $b=-0.01$ and show the adopted
conventions for the various lines. In particular, the gray dashed
[dotted] lines correspond to $n_{\rm s}=0.991$ [$n_{\rm
s}=0.935$], whereas the gray solid lines are obtained by fixing
$n_{\rm s}=0.963$ -- see Eq.~(\ref{nswmap}). The black solid lines
correspond to $M=M_{\rm GUT}$, whereas beyond the black dashed
lines the preferred hierarchy in the violation of
Eqs.~(\ref{flat}{a}) and (\ref{flat}{b}) fails. We observe that
the latter requirement holds only for $\kp\lesssim\ld$. Along the
gray solid line ($n_{\rm s}=0.963$), we obtain
$2.7\times10^{13}\lesssim T_{\rm
1rh}/\GeV\lesssim1.5\times10^{14}~$ and $1.9\times10^{-14}\lesssim
Y_{2\Gr}\lesssim10^{-13}$ for
$2\times10^{-4}\lesssim\kp\lesssim0.113$. Note that $T_{\rm
2rh}=3\times10^4~\GeV$ remains fixed since $\lambda_\mu$ and $f_a$
are also fixed throughout our computation.

One of the outstanding features of our proposal is that the
reduction of $\ns$ can be attained without disturbing the
monotonicity of the potential, contrary to other similar
suggestions -- see Refs.~\cite{hinova, gpp, king}. This fact can
be highlighted in Fig.~\ref{Vhi}, where we present the variation
of the inflationary potential $\Vhi$ as a function of the inflaton
field, $\sg$, for the values of the parameters corresponding to
the three intersections of the solid black line with the gray
lines in Fig.~\ref{kaM}(a) and (b). Namely, we take $M=M_{\rm
GUT}$, $b=-0.01$, and $\kappa=0.0036,~\ck=-0.0062$, and $\sg_{\rm
HIi}=s_{\rm HIi}=q_{\rm HIi}=3\times10^{17}~\GeV$ ($\ns=0.991$,
dashed line) or $\kappa=0.002,~\ck=-0.0125$, and $\sg_{\rm
HIi}=s_{\rm HIi}=q_{\rm HIi}=2.5\times10^{17}~\GeV$ ($\ns=0.963$,
solid line) or $\kappa=0.00111,~\ck=-0.0189$, and $\sg_{\rm
HIi}=s_{\rm HIi}=q_{\rm HIi}=2\times10^{17}~\GeV$ ($\ns=0.935$,
dotted line). The values corresponding to $\sigma_*$ and
$\sigma_{\rm f}$ are also designed. We observe that for large
values of $\sg$, $\Vhi$ develops an oscillatory behavior due to
the initial oscillations of $s$ and $q$ -- see Fig.~\ref{fN}.
However, $\Vhi$, for lower $\sg$'s, remains monotonic and,
therefore, no complications arise in the realization of the
inflationary dynamics.

\begin{figure}[!t]
\includegraphics[width=60mm,angle=-90]{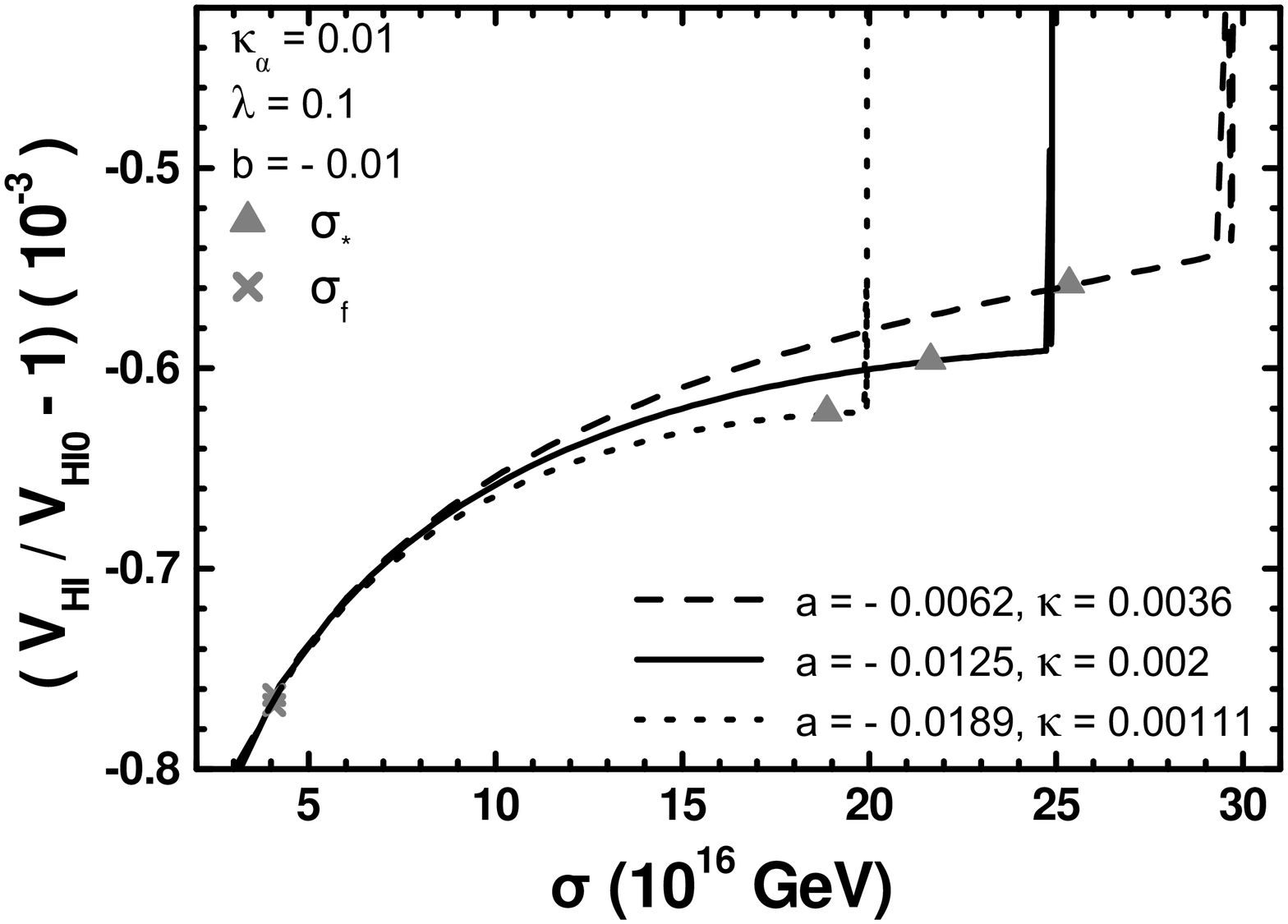}
\caption{The variation of the inflationary potential $\Vhi$ as a
function of $\sg$ for $M=M_{\rm GUT}$,
$\ka=-b=0.01,~\ld=c=d=e=f=g=0.1$, and
$\kappa=0.0036,~\ck=-0.0062,$ and $\sg_{\rm HIi}=s_{\rm
HIi}=q_{\rm HIi}=3\times10^{17}~\GeV$ ($\ns=0.991$, dashed line)
or $\kappa=0.002,~\ck=-0.0125,$ and $\sg_{\rm HIi}=s_{\rm
HIi}=q_{\rm HIi}=2.5\times10^{17}~\GeV$ ($\ns=0.963$, solid line)
or $\kappa=0.00111,~\ck=-0.0189,$ and $\sg_{\rm HIi}=s_{\rm
HIi}=q_{\rm HIi}=2\times10^{17}~\GeV$ ($\ns=0.935$, dotted line).
The values corresponding to $\sigma_*$ and $\sigma_{\rm f}$ are
also depicted.}\label{Vhi}
\end{figure}

Letting $b$ vary for a number of fixed values of $\ck$ and for the
most exciting case with $M=M_{\rm GUT}$, we can depict the values
allowed  by all the constraints of Sec.~\ref{cont} in the $\kp-b$
plane -- see Fig.~\ref{kb}. The various lines terminate at low
[high] $\kappa$'s due to the saturation of Eq.~(\ref{nswmap}) from
below [above]. The central $\ns$ is obtained at $\kp\simeq0.002$.
We readily conclude that the allowed $\ck$'s and $b$'s for
$M=M_{\rm GUT}$ and fixed $\ns$ are almost $\kp$-independent. This
is because, for fixed $\ns$, $\msm$ is fixed too. In particular,
for $\ns=0.963$, we have $\msm=-0.0136$, whereas, for
$\ns=0.935~[\ns=0.991]$, we have
$\msm\simeq-0.02~[\msm\simeq-0.008]$ and
$\kp\simeq0.0036~[\kp\simeq0.0011]$. In all cases,
$\msp\simeq1.82$. For this reason, we can present in Fig.~\ref{ab}
the allowed values by Eqs.~(\ref{Ntott}), (\ref{Prob}),  and
(\ref{nswmap}) in the $\ck-b$ plane for $M\simeq M_{\rm GUT}$ and
$\kp\simeq0.002$ ($\ns=0.963$, solid lines), $\kp\simeq0.0036$
($\ns=0.991$, dashed lines) or $\kp\simeq0.0011$ ($\ns=0.935$,
dotted lines). We observe that our scenario can be realized for
both signs of $\ck$ and $b$, contrary to the cases studied in
Refs.~\cite{hinova, king, gpp} where negative $b$'s are
necessitated. Moreover, compared to the latter cases, larger
$|b|$'s (of the order of $0.1$) are here permitted.

\begin{figure}[!b]
\centering\includegraphics[width=60mm,angle=-90]{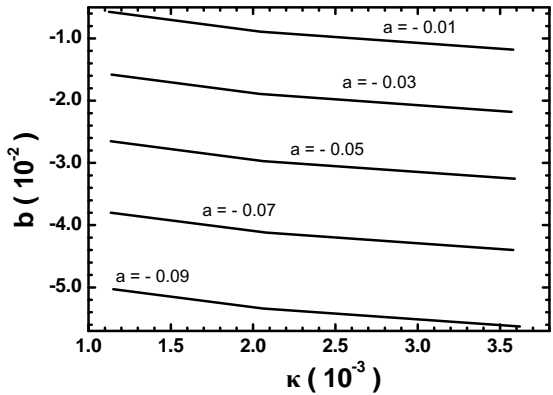}
\caption{\label{kb} Allowed values by Eqs.~(\ref{Ntott}),
(\ref{Prob}), and (\ref{nswmap}) in the $\kappa-b$ plane for
$\ka=0.01,~\ld=c=d=e=f=g=0.1,~M=M_{\rm GUT},$ and various $\ck$'s
indicated on the curves.}
\end{figure}

%Let us finally add that the desired above $\msm$'s and $\msp$'s
%can be derived even with positive $\ck$'s and $b$'s. E.g., for
%$\ck=0.02$ we get $b=0.0085,0.0049$ and $0.002$ for
%$\msm\simeq-0.02,-0.0136$ and $-0.008$ correspondingly.

\begin{figure}[t!]
\centering\includegraphics[width=60mm,angle=-90]{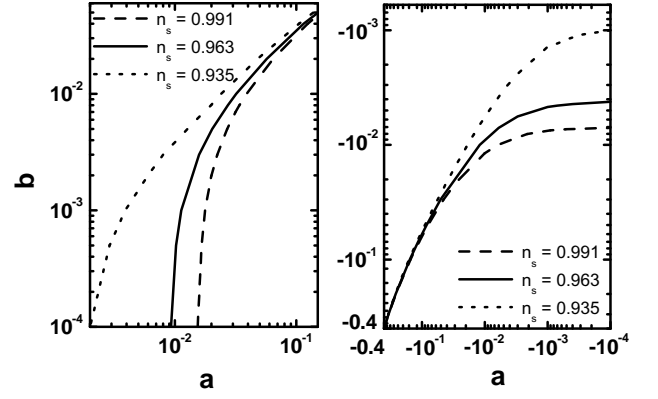}
\caption{\label{ab} Allowed values by Eqs.~(\ref{Ntott}),
(\ref{Prob}), and (\ref{nswmap}) in the $\ck-b$ plane for
$\ka=0.01,~\ld=c=d=e=f=g=0.1,~M\simeq M_{\rm GUT},$ and
$\kp\simeq0.002$ (solid lines), $\kp\simeq0.0036$ (dashed lines)
or $\kp\simeq0.0011$ (dotted lines).}
\end{figure}

\section{Conclusions \label{con}}  We investigated a cosmological
scenario according to which a SUSY GUT scale FHI is followed by a
PQPT which resolves the strong CP and the $\mu$ problems of MSSM.
The PQPT is tied to renormalizable superpotential terms and the
possible catastrophic production of domain walls can be eluded by
the introduction of extra matter superfields which can be chosen
so that the MSSM gauge coupling constant unification is not
disturbed. The inflaton-like field, associated with PQPT, plays a
crucial role in the construction of the \Ka\ which is expanded up
to fourth order in powers of the various fields. The FHI
reproduces the current data on $P_{\cal R}$ and $n_{\rm s}$ within
the power-law $\Lambda$CDM cosmological model and generates the
number of e-folds required from the resolution of the horizon and
flatness problems of the SBB. The dynamics of FHI is investigated
both numerically and analytically and the results are compared
with each other. Fixing $n_{\rm s}$ to its central value and $M$
to the SUSY GUT scale, we concluded that $\kp=0.002$ with the
remaining parameters taking more or less natural values,
$\pm(0.01-0.1)$. It is gratifying that our model supports a second
stage of reheating after PQPT, which dilutes sufficiently the
$\Gr$ abundance so as to become observationally safe for $\Gr$
masses even lower than $10~{\rm TeV}$.

\acknowledgments We would like to thank I. Moss, R.O.~Ramos, V.N.
\c{S}eno\u{g}uz, and Q.~Shafi for useful discussions. This work
was supported by the European Union under the Marie Curie Initial
Training Network ``UNILHC'' PITN-GA-2009-237920 and the Marie
Curie Research Training Network ``UniverseNet''
MRTN-CT-2006-035863.

\appendix

\section{Reheating Processes and Gravitino Constraint}\label{Rhg}

In this Appendix, we present a numerical description of the
post-inflationary evolution of the various energy densities in our
set-up paying special attention to the dilution of the $\Gr$
yield, $Y_{\Gr}$.

The energy density, $\rho_1$ [$\rho_2$], of the oscillatory system
which reheats the universe at the temperature $T_{1\rm rh}$
[$T_{2\rm rh}$], the energy density of produced radiation,
$\rho_{\rm R}$, and the number density of $\Gr$, $n_{\Gr}$,
satisfy the following Boltzmann equations -- cf. Refs.~\cite{gpp,
kohri}:
\begin{eqnarray}
&& \dot \rho_1+3H\rho_1+\Gamma_1
\rho_1=0,\label{nf}\\
&& \dot\rho_2+3H\rho_2+\Gamma_2\rho_2=0,\label{nfb} \\
&& \dot\rho_{\rm R}+4H\rho_{\rm R}-
\Gamma_1\rho_1-\Gamma_2\rho_2=0,\label{rR}\\
&& \dot n_{\Gr}+3Hn_{\Gr}-C_{\Gr} \lf n^{\rm eq}\rg^2=0.\label{ng}
\end{eqnarray}
Here $n^{\rm eq}={\zeta(3)T^3/\pi^2}$ is the equilibrium number
density of the bosonic relativistic species, $C_{\Gr}$ is a
collision term for $\Gr$ production which, in the limit of the
massless gauginos, turns out to be \cite{kohri, brand}
\beq C_{\Gr} = \frac{3\pi}{16\zeta(3)\mP^2}\sum_{i=1}^{3} c_i
g_i^2 \ln\left({k_{i}\over g_i}\right),\eeq
where $g_i$ (with $i=1,2,3$) are the gauge coupling constants of
MSSM, $(k_i)=(1.634,1.312,1.271),$ and $(c_i)=(33/5,27,72)$. Also
the Hubble expansion parameter, $H$, during this period is given
by
\begin{equation} \label{Hini}
H=\frac{1}{\sqrt{3}\mP} \left(m_{{\Gr}}n_{{\Gr}}+\rho_1
+\rho_2+\rho_{\rm R} \right)^{1/2}.
\end{equation}
Clearly, in the limit of massless MSSM gauginos, the $n_{\Gr}$
computation is $m_{\Gr}$-independent. The temperature, $T$, and
the entropy density, ${\sf s}$ (not to be confused with the field
$s$), can be found using the relations
\begin{equation} \rho_{\rm R}=\frac{\pi^2}{30}g_*
T^4~\mbox{and}~{\sf s}=\frac{2\pi^2}{45}g_* T^3,
\label{rs}\end{equation}
where $g_* (T)=g_{1\rm rh*}$ [$g_* (T)=g_{2\rm rh*}$] for $T\geq
T_{\rm PQ}$ [$T<T_{\rm PQ}$] with $T_{\rm PQ}$ being defined as
the solution of the equation $\rho_{\rm R}\lf T_{\rm PQ}\rg=\Vpqo$
and can be found numerically.

The numerical integration of Eqs.~(\ref{nf})--(\ref{ng}) is
facilitated by absorbing the dilution terms. To this end, we find
it convenient to define \cite{quin} the following dimensionless
variables
\begin{equation} \label{fdef}
f_1=\rho_1 R^3,~f_2=\rho_2 R^3,~f_{\rm R}=\rho_{\rm R}
R^4,~\mbox{and}~f_{{\Gr}}=n_{{\Gr}} R^3.
\end{equation}
Converting the time derivatives to derivatives w.r.t $\bar
N=\ln\left(R/R_{\rm HIf}\right)$ with $R_{\rm HIf}$ being the
value of the scale factor at the end of FHI (the value of $R_{\rm
HIf}$ turns out to be numerically irrelevant),
Eqs.~(\ref{nf})--(\ref{ng}) become
\begin{eqnarray}
Hf^\prime_1&=&-\Gamma_1 f_1,\label{ff}\\
Hf^\prime_2&=&-\Gamma_2 f_2,\label{ffb}\\
Hf^\prime_{\rm R}&=&\Gamma_1 f_1 R+\Gamma_2 f_2 R, \label{fR}\\
Hf^\prime_{\Gr}&=&C_{\Gr}\lf n^{\rm eq}\rg^2R^3,\label{fg}
\end{eqnarray}
where the differentiation w.r.t. $\bar N$ and $N$ -- see
Sec.~\ref{fhi1dyn} -- coincides. Also $H$ and $T$ can be expressed
in terms of the variables in Eq.~(\ref{fdef}) as
\beq \label{H2exp}  H=\frac{\sqrt{m_{\Gr}f_{\Gr}+f_1 +f_2 +f_{\rm
R}/R}}{\sqrt{3R^3}\mP}~\mbox{and}~ T=\sqrt[4]{{30\ f_{\rm
R}\over\pi^2 g_* R^4}}\cdot\eeq
The system of Eqs.~(\ref{ff})--(\ref{fg}) can be solved, imposing
the following initial conditions (the quantities below are
considered functions of the independent variable $\bar N$):
\beq\rho_1(0)=\Vhio,~\rho_{\rm R}(0)=n_{\Gr}(0)=0,~\mbox{and}~
\rho_2(\bar N_{\rm PQ})=\Vpqo, \label{init} \eeq
where $\bar N_{\rm PQ}$ is the value of $\bar N$ corresponding to
the temperature $T_{\rm PQ}$. Needless to say that we set
$\rho_2(\bar N)=0$ for $\bar N<\bar N_{\rm PQ}$.

\begin{figure}[!t]
\centering\includegraphics[width=60mm,angle=-90]{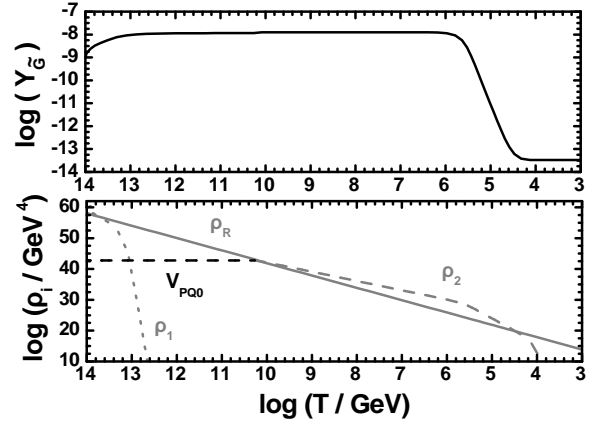}
\caption{\label{Trg} The evolution of the quantities $\log\rho_i$
with $i=1$ (gray dotted line), $i=2$ (gray dashed line), $i={\rm
R}$ (gray line), $\log\Vpqo$ (black dashed line), and $\log
Y_{\Gr}$ (black solid line) as functions of $\log T$ for
$\kappa=0.002, \ka=-b=0.01, \ck=-0.0125, \ld=c=d=e=f=g=0.1,
M=M_{\rm GUT}, f_a=10^{12}~\GeV$, and $\ld_\mu=0.01$.}
\end{figure}

In Fig.~\ref{Trg}, we illustrate the cosmological evolution of the
quantities $\log\rho_i$ with $i=1$ (dotted gray line), $i=2$
(dashed gray line), and $i={\rm R}$ (gray line), $\log\Vpqo$
(black dashed line), and $\log Y_{\Gr}$ (black solid line) as
functions of $\log T$ for the values of the parameters adopted in
Fig.~\ref{fN}. In particular, the parameters which determine the
evolution of the various quantities during the post-inflationary
period are $\kappa, \ld, M, \ka, f_a,$ and $\ld_\mu$. We take
$\kappa=0.002, \ld=0.1, M=M_{\rm GUT}, \ka=0.01,
f_a=10^{12}~\GeV,$ and $\ld_\mu=0.01$. From Fig.~\ref{Trg}, we
observe that FHI is followed successively by the following four
epochs: (i) a MD era, due to the oscillating and decaying inflaton
system, which lasts until $T\simeq T_{1\rm rh}$ given by
Eq.~(\ref{T1rh}), (ii) a RD epoch, which terminates at $T_{\rm
PQ}$, (iii) a MD era created by the oscillations of the PQ-system,
which is completed at $T\simeq T_{2\rm rh}$ given by
Eq.~(\ref{T2rh}), and (iv) a RD epoch after which the universe
enters the SBB phase. The completion of the two reheating
processes corresponds to the two intersections of $\rho_1$ and
$\rho_2$ with $\rho_{\rm R}$ in Fig.~\ref{Trg}. Here we  omit for
simplicity any possible instantaneous domination of $\Vpqo$ over
the various energy densities since, as we mention in
Sec.~\ref{fhi2}, $\Vpq$ does not support any significant period of
inflation.  In Fig.~\ref{Trg}, we also see that the ${\Gr}$ yield,
$Y_{\Gr}=n_{{\Gr}}/{\sf s}$, takes the value
$Y_{1\Gr}\simeq1.1\times10^{-8}$ -- see Eq.~(\ref{Y1}) -- for
$T\simeq T_{1\rm rh}$. However, due to the entropy released during
the out-of-equilibrium decay of the PQ system, $Y_{\Gr}$ decreases
sharply to $Y_{2\Gr}\simeq3.3\times10^{-14}$ -- see
Eq.~(\ref{Y2}). This result can be understood from the following
relation

\vspace*{-0.35cm} \beq Y_{2\Gr}=Y_{1\Gr}\frac{{\sf s}\lf T_{\rm
PQ}\rg}{{\sf s}\lf T_{\rm 2rh}\rg}\lf\frac{R_{\rm PQ}}{R_{\rm
2rh}}\rg^3, \label{Y12} \eeq
where $R_{\rm PQ}~[R_{\rm 2rh}]$ is the value of the scale factor
corresponding to $T_{\rm PQ}~[T_{\rm 2rh}]$. Taking into account
Eq.~(\ref{rs}) and the fact that, during this MD era,
$R\propto\rho_{\rm osc}^{-1/3}$ -- recall that $\rho_{\rm osc}$ is
the energy density of the oscillating system -- we arrive at
Eq.~(\ref{Y2}). We verify that the analytical results agree
remarkably with the numerical ones.

Therefore, we can easily appreciate the importance of PQPT, as
realized in our scenario, in lowering $Y_{\Gr}$ to a value
compatible with the observational data \cite{kohri}.

\def\ijmp#1#2#3{{Int. Jour. Mod. Phys.}
{\bf #1},~#3~(#2)}
\def\plb#1#2#3{{Phys. Lett. B }{\bf #1},~#3~(#2)}
\def\zpc#1#2#3{{Z. Phys. C }{\bf #1},~#3~(#2)}
\def\prl#1#2#3{{Phys. Rev. Lett.}
{\bf #1},~#3~(#2)}
\def\rmp#1#2#3{{Rev. Mod. Phys.}
{\bf #1},~#3~(#2)}
\def\prep#1#2#3{{Phys. Rep. }{\bf #1},~#3~(#2)}
\def\prd#1#2#3{{Phys. Rev. D }{\bf #1},~#3~(#2)}
\def\npb#1#2#3{{Nucl. Phys. }{\bf B#1},~#3~(#2)}
\def\npps#1#2#3{{Nucl. Phys. B (Proc. Sup.)}
{\bf #1},~#3~(#2)}
\def\mpl#1#2#3{{Mod. Phys. Lett.}
{\bf #1},~#3~(#2)}
\def\arnps#1#2#3{{Annu. Rev. Nucl. Part. Sci.}
{\bf #1},~#3~(#2)}
\def\sjnp#1#2#3{{Sov. J. Nucl. Phys.}
{\bf #1},~#3~(#2)}
\def\jetp#1#2#3{{JETP Lett. }{\bf #1},~#3~(#2)}
\def\app#1#2#3{{Acta Phys. Polon.}
{\bf #1},~#3~(#2)}
\def\rnc#1#2#3{{Riv. Nuovo Cim.}
{\bf #1},~#3~(#2)}
\def\ap#1#2#3{{Ann. Phys. }{\bf #1},~#3~(#2)}
\def\ptp#1#2#3{{Prog. Theor. Phys.}
{\bf #1},~#3~(#2)}
\def\apjl#1#2#3{{Astrophys. J. Lett.}
{\bf #1},~#3~(#2)}
\def\n#1#2#3{{Nature }{\bf #1},~#3~(#2)}
\def\apj#1#2#3{{Astrophys. J.}
{\bf #1},~#3~(#2)}
\def\anj#1#2#3{{Astron. J. }{\bf #1},~#3~(#2)}
\def\mnras#1#2#3{{MNRAS }{\bf #1},~#3~(#2)}
\def\grg#1#2#3{{Gen. Rel. Grav.}
{\bf #1},~#3~(#2)}
\def\s#1#2#3{{Science }{\bf #1},~#3~(#2)}
\def\baas#1#2#3{{Bull. Am. Astron. Soc.}
{\bf #1},~#3~(#2)}
\def\ibid#1#2#3{{\it ibid. }{\bf #1},~#3~(#2)}
\def\cpc#1#2#3{{Comput. Phys. Commun.}
{\bf #1},~#3~(#2)}
\def\astp#1#2#3{{Astropart. Phys.}
{\bf #1},~#3~(#2)}
\def\epjc#1#2#3{{Eur. Phys. J. C}
{\bf #1},~#3~(#2)}
\def\nima#1#2#3{{Nucl. Instrum. Meth. A}
{\bf #1},~#3~(#2)}
\def\jhep#1#2#3{{J. High Energy Phys.}
{\bf #1},~#3~(#2)}

\end{document}